%% file: main.tex
\documentclass[acmsmall]{acmart}

\AtBeginDocument{%
  }

\setcopyright{acmlicensed}
\copyrightyear{2025}
\acmYear{2025}
\acmDOI{XXXXXXX.XXXXXXX}

\acmSubmissionID{2551}

\usepackage{color}
\usepackage{soul}
\usepackage{graphicx} 
\usepackage{subcaption} 
\usepackage{graphicx}
\usepackage{tabularray}
\usepackage{array, blindtext}
\usepackage{booktabs}
\usepackage{hyperref}
\usepackage{needspace}

\begin{document}

\title{Exploring Collaborative GenAI Agents in Synchronous Group Settings: Eliciting Team Perceptions and Design Considerations for the Future of Work}
\renewcommand{\shorttitle}{Exploring Collaborative Agents in Synchronous Group Settings}



\author{Janet G. Johnson}
\affiliation{%
 \institution{University of Michigan}
 \city{Ann Arbor}
 \state{Michigan}
 \country{USA}}

\author{Macarena Peralta}
\affiliation{%
 \institution{University of Michigan}
 \city{Ann Arbor}
 \state{Michigan}
 \country{USA}}

\author{Mansanjam Kaur}
\affiliation{%
 \institution{University of Michigan}
 \city{Ann Arbor}
 \state{Michigan}
 \country{USA}}

\author{Ruijie Sophia Huang}
\affiliation{%
 \institution{University of Michigan}
 \city{Ann Arbor}
 \state{Michigan}
 \country{USA}}

\author{Sheng Zhao}
\authornote{Author contributed to this research during an internship at the University of Michigan, Ann Arbor, USA.}
\affiliation{%
 \institution{Tsinghua University}
 \city{Beijing}
 \country{China}}

\author{Ruijia Guan}
\affiliation{%
 \institution{University of Michigan}
 \city{Ann Arbor}
 \state{Michigan}
 \country{USA}}

\author{Shwetha Rajaram}
\affiliation{%
 \institution{University of Michigan}
 \city{Ann Arbor}
 \state{Michigan}
 \country{USA}}

\author{Michael Nebeling}
\affiliation{%
 \institution{University of Michigan}
 \city{Ann Arbor}
 \state{Michigan}
 \country{USA}}

\renewcommand{\shortauthors}{Johnson et al.}

\input{0_Abstract}

\ccsdesc[500]{Human-centered computing~Empirical studies in collaborative and social computing}
\ccsdesc[500]{Human-centered computing~Mixed / augmented reality}

\keywords{Collaborative Problem-Solving, Generative Aritifical Intelligence, Mixed Reality, Synchronous Groupwork}


\maketitle

\begin{center}
\begin{minipage}{\linewidth} 
\textit{This is a pre-print of an article accepted at the ACM SIGCHI Conference on Computer-Supported Cooperative Work and Social Computing (CSCW 2025).}
\end{minipage}
\end{center}

\input{1_Introduction}
\input{2_Background}
\input{3_Provotype}

\input{4_Method}
\input{5_Findings1}
\input{6_Findings2}

\input{7_Findings3}
\input{8_Takeaways}
\input{9_Discussion}

\input{10_Conclusion}




\bibliographystyle{ACM-Reference-Format}
\bibliography{1_references}

\input{11_Appendix}


\end{document}

%% file: 0_Abstract.tex
\begin{abstract}
While generative artificial intelligence (GenAI) is finding increased adoption in workplaces, current tools are primarily designed for individual use. Prior work established the potential for these tools to enhance personal creativity and productivity towards shared goals; however, we don't know yet how to best take into account the nuances of group work and team dynamics when deploying GenAI in work settings. In this paper, we investigate the potential of \textit{collaborative} GenAI agents to augment teamwork in synchronous group settings through an exploratory study that engaged 25 professionals across 6 teams in speculative design workshops and individual follow-up interviews. Our workshops included a mixed reality \textit{provotype} to simulate embodied collaborative GenAI agents capable of actively participating in group discussions. Our findings suggest that, if designed well, collaborative GenAI agents offer valuable opportunities to enhance team problem-solving by challenging groupthink, bridging communication gaps, and reducing social friction. However, teams' willingness to integrate GenAI agents depended on its perceived fit across a number of individual, team, and organizational factors. We outline the key design tensions around agent representation, social prominence, and engagement and highlight the opportunities spatial and immersive technologies could offer to modulate GenAI influence on team outcomes and strike a balance between augmentation and agency.
\end{abstract}

%% file: 1_Introduction.tex
\section{Introduction}

Generative artificial intelligence (GenAI) -- with its capacity to create seemingly novel, human-level artifacts -- has shown potential to support creative-problem solving and augment our collective capabilities~\cite{shi2023understanding,fui2023generative, morris2023design}. The increased adoption of GenAI tools within teams and in the workplace makes it inevitable that it will fundamentally transform the future of work~\cite{woodruff2024knowledge,cazzaniga2024gen,eloundou2024gpts}. Its advanced reasoning capabilities that can support both divergent and convergent aspects of problem-solving~\cite{inie2023designing, wang2024exploring}, and its ability to simulate social behavior and dynamically adapt to evolving contextual cues~\cite{muller2024genaichi,wang2024survey,park2022social} could make GenAI especially well-suited to engage in the inherently collaborative analytical and dialogic aspects of problem-solving that help teams tackle complexity in ways that surpasses what individuals might achieve alone~\cite{cui2024ai, malone2018superminds, wuchty2007increasing}.

However, today's GenAI tools are primarily designed for individuals, and while this may enhance individual productivity and contributions to shared goals, they overlook how problem-solving unfolds within teams~\cite{malone2018superminds, wuchty2007increasing}. These tools are also fundamentally reactive and rely on user-generated prompts and add an additional overhead that detracts from the shared workspace and disrupts the flow of the collaboration in group settings~\cite{van2024exploring,han2024teams}. 

While prior work suggests that there is a growing interest in proactive AI agents that can function as a peer-like collaborator ~\cite{hwang2021ideabot,zhou2024understanding,salikutluk2024evaluation}, research around GenAI agents in group settings remain scarce~\cite{siemon2022elaborating,van2024exploring}. \textbf{In this paper, we explore this untapped potential for \textit{collaborative} GenAI agents to engage with teams in synchronous group settings and become an active collaborator that can independently contribute towards collective goals}. We aim to provide a balanced perspective by exploring both the potential benefits and the concerns of a future where GenAI agents actively engage in collaborative problem-solving.

\begin{figure} [b]
  \centering
  \includegraphics[width=\linewidth]{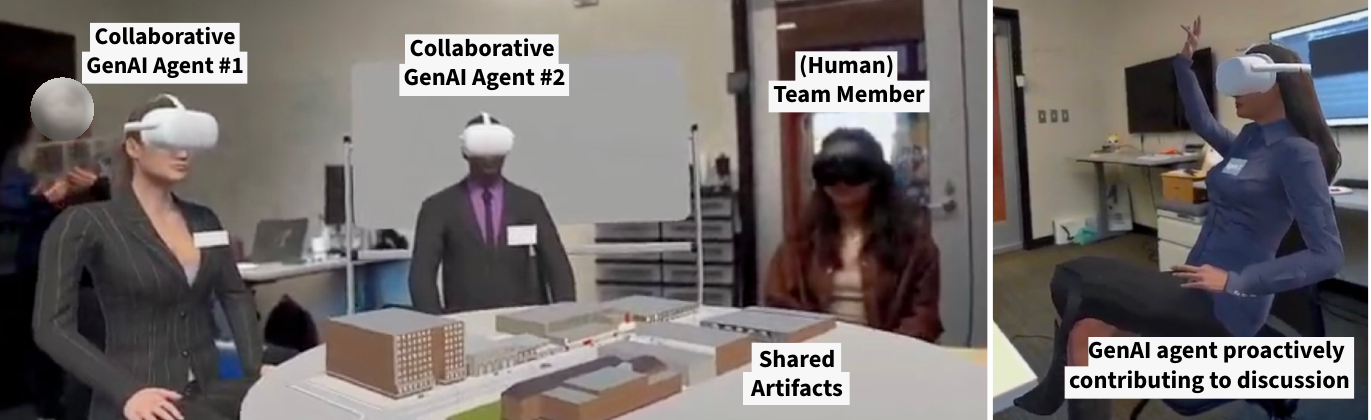}
  \caption{Our Mixed Reality provotype showcased one possible future where teams share their physical space with collaborative GenAI agents (Left) that could actively participate in a deliberative group discussion (Right).}
  \Description{}
  \label{fig:provotype}
\end{figure}

Our investigation addresses the many unanswered questions around the desirability and perceived value associated with integrating GenAI agents into team settings~\cite{seymour2024speculating}, as well as their impact on team dynamics and group outcomes~\cite{seeber2020machines,he2024ai,shin2023integrating}. Moreover, teams operate as a cohesive entity that extends beyond individual members and are often situated within larger organizational ecosystems -- this means that addressing broader team and organizational requirements in addition to the individual considerations is crucial, as mistakes and setbacks can lead to costly failures~\cite{flathmann2020invoking}. Understanding end-user needs along individual, team, and organizational aspects is therefore essential to designing a collectively desired future~\cite{rheingold1985tools,mitrovic2021beyond} and bridging the socio-technical gap to ensure that technology supports, rather than dictates, the future of work~\cite{ackerman2000intellectual}.

Additionally, collaboration in synchronous group settings typically involves multiple people engaging through verbal and non-verbal communication in an embodied and situated manner \cite{doherty1997face, hughes1997cooperative, harrison1996re}. Integrating GenAI as an active collaborator in teams will therefore require moving beyond the current trend of screen-based chatbot interactions~\cite{van2024exploring}. Spatial and immersive technologies like mixed reality (MR) enable rich multimodal and embodied interactions that are especially effective in group settings as they allow users to engage with both digital content and their collaborators in relatively natural and intuitive ways~\cite{billinghurst2002collaborative,speicher2019mixed}. Early investigations signal that both voice-based and embodied conversational agents may enhance human-AI interactions~\cite{billinghurst2024social, mahmood2023llm, bovo2024embardiment} and offer a compelling alternative to screenbound agent interfaces. 

However, we don't currently understand what, if any, opportunities this may bring to team interactions with shared collaborative GenAI agents. Therefore, in this paper, we also explore what the additional spatial and immersive capabilities of MR bring to GenAI-enhanced synchronous group settings. Although today's MR devices have their limitations, MR as a medium integrates many technologies and interaction modalities that enable a wide range of user experiences~\cite{speicher2019mixed,skarbez2021revisiting}. Leveraging the flexibility and expressiveness of MR in our explorations would also allow end-users to imagine a future unconstrained by their notions of technological limitations, which in turn allows us to better understand how collaborative GenAI systems can meet end-user needs.

Our work in this paper is therefore guided by the following three research questions: 
\begin{itemize}
    \item [\textbf{RQ1}] What value do end-users expect collaborative GenAI agents to provide in enhancing team capabilities in synchronous group settings?
    \item [\textbf{RQ2}] What anticipated needs and challenges at individual, team, and organizational levels inform design considerations and influence the acceptance of GenAI agents within teams?
    \item [\textbf{RQ3}] What additional opportunities do more expressive technologies like mixed reality afford for meeting group needs when interacting with collaborative GenAI agents?
\end{itemize}

\begin{figure} [b]
  \centering
  \includegraphics[width=\linewidth]{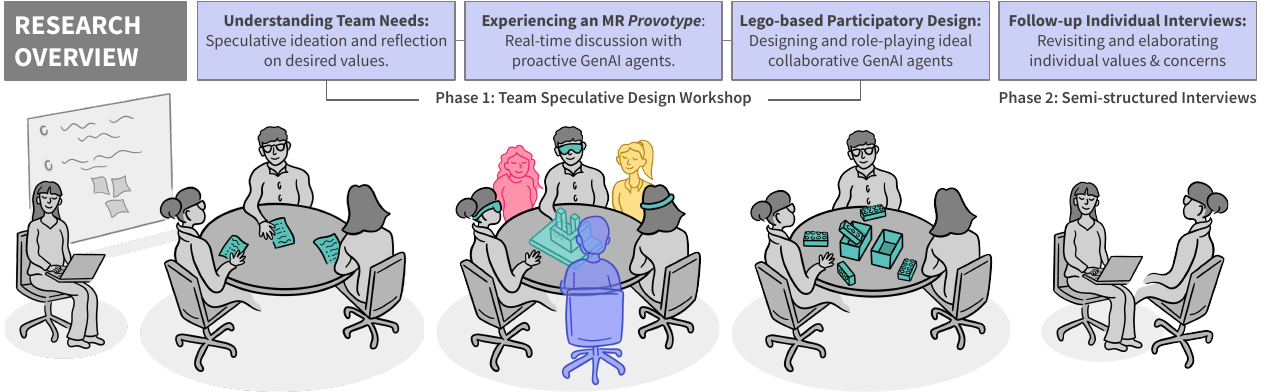}
  \caption{ An overview of our methods that included a speculative design workshop as well as individual follow-up interviews with teams. The team workshops employed a mix of methods including speculative ideation and reflection, embodied interactions with GenAI agents through an MR provotype, and a participatory design activity where participants used Lego blocks to depict their ideal GenAI-based collaborative agents.}
  \Description{}
  \label{fig:overview}
\end{figure}

Our research centers on an exploratory study with 25 professionals in 6 teams across a variety of domains through speculative team workshops and individual semi-structured interviews (Fig.~\ref{fig:overview}). In addition to reflective and participatory activities, our workshops utilized \textit{provotypes} -- provocative prototypes intended to inspire discussion -- to illustrate a possible future where embodied collaborative GenAI agents in MR actively participated in group discussions (Fig.~\ref{fig:provotype}). This allowed us to uncover design considerations and future research directions that aren't constrained by the current trend of GenAI interfaces. Engaging directly with real teams also allowed us to capture authentic perspectives from those who would ultimately interact with collaborative GenAI agents, grounding our findings in the lived experiences, values, and practical needs of future end-users. 

Overall, \textbf{our participants perceived significant value in collaborative GenAI agents} and predicted that they could challenge groupthink, reduce social friction, and boost confidence in team outcomes. However, they also raised a number of concerns and stressed that \textbf{their willingness to adopt collaborative GenAI agents depended on its perceived fit with both individual and team needs, as well as the presence of organizational structures and guardrails that safeguard collective human interests}. Key considerations included being able to trust and evaluate a collaborative agent's usefulness, gauging its cultural alignment, and mitigating negative impacts on their relationships, processes, and outcomes. Participants also stressed the importance of psychological safety in work environments and advocated for human-centered guidelines and thoughtful deployment of GenAI within teams. Our study also uncovered key design tensions around a collaborative GenAI agent's influence on groups and the balance between collective evaluation and individual agency. Finally, we outline the future research requirements to ensure appropriate reliance on GenAI within group settings, and \textbf{highlight a major opportunity for spatial mediums like MR to function as a modulator through which GenAI's influence on teams can be managed or regulated}.

%% file: 2_Background.tex
\section{Background and Related Work}

\subsection{GenAI for Problem-Solving in Synchronous Group Settings}

GenAI has shown significant value in enhancing our creativity and ability to engage in the different aspects of problem-solving~\cite{hubert2024current, wadinambiarachchi2024effects} including ideation~\cite{liu2024personaflow, heyman2024supermind}, problem-identification~\cite{jeon2021fashionq}, prototyping~\cite{epstein2022happy,lc2023speculative}, deliberation~\cite{chiang2024enhancing}, and decision making~\cite{ma2024towards}. However, most of the work in the field focuses on exploring how GenAI tools share agency with a single user to complete a task. To fill this gap, researchers have explored how users can use GenAI tools individually while simultaneously interacting with other users during collaborative design and ideation~\cite{koch2020imagesense,deng2024crossgai}, qualitative analysis~\cite{gao2023coaicoder}, and during a cooperative game~\cite{sidji2024human}. Recent work has also explored how multiple users can engage with either ChatGPT or custom-built LLM tools to enhance co-located brainwriting~\cite{shaer2024ai} and group ideation \cite{he2024ai}. For example, Suh et al. investigated how collaboratively composing music with a GenAI tool impacted social dynamics~\cite{suh2021ai} and Zhang et al. evaluated how an LLM-driven shared display could support ideation and discussion within colocated teams~\cite{zhang2024ladica}.

While these initial explorations with GenAI for groupwork show promise in enhancing team outcomes, these interventions primarily operate more as supportive "AI-infused supertools"~\cite{shneiderman2022human} rather than actively contributing to the collaborative goals. In fact, some of the findings in these studies suggest a desire for AI agents to be more proactive~\cite{koch2020imagesense, he2024ai}. For instance, Hwang et al. found that participants produced more and higher-quality ideas when they perceived their AI partner to be an autonomous agent that could act as an active teammate~\cite{hwang2021ideabot}. Similarly, Salikutluk et al. demonstrated that users engaged in cooperative tasks preferred AI agents with higher autonomy levels~\cite{salikutluk2024evaluation}, and Zhou et al. observed that participants preferred intelligent agents to be a peer-like collaborator in creative design contexts~\cite{zhou2024understanding}. This preference is reinforced by Zhang et al., who found that proactive communication from AI enhanced trust and situational awareness~\cite{zhang2023investigating}.

Moreover, these applications are fundamentally reactive and requires users to explicitly invoke or prompt the system for it to generate content. While this offers a low barrier to access creative possibilities, non-AI professionals often find it difficult to express their intentions and effectively achieve their desired results with GenAI~\cite{zamfirescu2023johnny, xiaohan2024designing, wadinambiarachchi2024effects}. These interactions also require users to divide their attention between prompting the system and the collaborative task at hand, which inadvertently detracts from the shared workspace~\cite{van2024exploring} and disrupts the flow of group discussions~\cite{han2024teams}. This increased cognitive overhead can introduce friction and a "crooked bowtie" effect~\cite{verheijden2023collaborative} that reduces the overall efficiency of group interactions. 

\subsection{Collaborative GenAI Agents and Understanding User Needs}

Autonomously creative agents are artificially intelligent entities capable of independently perceiving, processing, and responding to their environment in a timely manner~\cite{russell2016artificial,jennings2010developing} and can operate independently without the need for constant human input~\cite{salikutluk2024evaluation}. The emerging capabilities of GenAI allow AI agents to directly participate in social interaction~\cite{muller2024genaichi}, and there is growing evidence that LLMs in particular can simulate human behaviors and cognitive processes~\cite{park2023generative,park2022social,wang2024survey}. There is, therefore, an untapped potential for GenAI-based collaborative agents to take on more proactive roles where they dynamically interact with human or AI collaborators and independently offer novel contributions in group settings. These agents could be seen as genuine teammates capable of cognitively complex tasks, autonomous contributions, and reciprocal collaboration.

Few studies have integrated collaborative agents in group settings, though some progress has been made~\cite{siemon2022elaborating,van2024exploring}. For example, Chiang et al. explored an LLM-powered “devil’s advocate” that contributed to group deliberations~\cite{chiang2024enhancing}, Imamura et al. developed a system that displayed relevant articles on a wall display to enhance group discussions~\cite{imamura2024serendipity}, Hwang et al. explored the ability of LLM-based agents to represent a user in social settings~\cite{hwang2024whose}, and Samadi et al. built a GPT-based collaborator agent that could engage with a team over digital chat-based platforms like Slack~\cite{samadi2024ai}. While these explorations are promising, there are a lot of unanswered questions about what it means to design effective collaborative GenAI agents for teams~\cite{seeber2020machines}. There are also several risks associated with the possibility of it unintentionally introducing bias, disrupting team dynamics, interfering with power dynamics, or stifling non-conformity~\cite{seymour2024speculating}. There is, therefore, a need to understand the desirability, value, and ethical considerations associated with integrating autonomous collaborative agents in teams~\cite{he2024ai,koch2020imagesense,shin2023integrating}.

Designing GenAI interfaces that appropriately cater to human needs is a challenging HCI problem and one that has received a lot of attention in recent years~\cite{2023arXivTerryInteractive}. We've seen the development of a large number of UX design principles~\cite{weisz2024design} and conceptual frameworks that tackle the bi-directional alignment GenAI systems necessitate~\cite{shen2024towards}. When it comes to collaborating with AI agents, researchers have explored the team dynamics, communication strategies, and how personal characteristics shape user perceptions of AI~\cite{zhang2023investigating,flathmann2024empirically,tolzin2023mechanisms,zhang2024know}. However, with the exception of Flathmann et al.'s framework that takes into account multiple users~\cite{flathmann2020invoking}, most of the work focuses on Human-AI dyads. Several studies have also examined end-user perspectives on GenAI and found that people see GenAI intensifying social issues like deskilling, dehumanization, and disinformation~\cite{woodruff2024knowledge,inie2023designing}. Wang et al., in particular, looked at how GenAI impacts group dynamics within UX teams but primarily focused on the individual use of GenAI tools available today~\cite{wang2024exploring}. There is, therefore, a need to better understand end-users perspectives about future GenAI agents that work collectively with the whole team.

\subsection{Expressive Interfaces and Opportunities with MR}

The widespread adoption of applications like ChatGPT, Dall-E, Midjourney, and more has popularized engaging with GenAI through text and chatbot-like interfaces. While this has allowed people to explore the capabilities of LLMs and other GenAI technologies, recent research indicates that the design of these interfaces significantly impacts how effectively users can utilize these tools~\cite{heyman2024supermind, subramonyam2024bridging}. However, human collaboration is often multimodal, embodied, and situated as it involves a rich tapestry of communication methods that rely on multiple sensory channels where participants use both verbal and non-verbal forms of communication and engage with shared objects and spaces ~\cite{doherty1997face, hughes1997cooperative, kirsh2013embodied, harrison1996re}. Moreover, cognition is inherently spatial and our ability to think is often facilitated by our interaction with the physical space around us~\cite{tversky2015cognitive,kirsh2013embodied}. Considering collaborative mediums that go beyond screen-based interactions and traverse voice, gestures, and interactions with the environment could therefore be integral to integrating collaborative GenAI agents within teamwork settings~\cite{van2024exploring,seymour2024speculating}.

Immersive technologies like MR offer a compelling alternative to screenbound interfaces as their ability to overlay digital information onto the physical world reduces the separation between task and communication spaces~\cite{billinghurst2002collaborative}. Additionally, their inherently spatial nature enables more embodied and situated interactions that mimic real-world interactions~\cite{billinghurst2024social, bozkir2024embedding, chang2024emiras} while also allowing the exploration of new social interactions that go beyond traditional mechanisms~\cite{johnson2023unmapped,gronbaek2022designing}. Research on intelligent virtual agents shows that imbuing an agent with a visual body and social behaviors could provide users with additional communication cues~\cite{ReinhardtCAINAR2020} and increase their sense of engagement and lower task load in collaborative settings~\cite{kim2018ISMAREmbody,KimVR20}. Early investigations with voice-based GenAI agents~\cite{mahmood2023llm} and LLM-based embodied conversational agents~\cite{WanLLMBasedAgentCHI23,aikawa2023introducing,bovo2024embardiment} also demonstrate how embodied or conversational agents could enhance human-AI interactions and expand the impact of LLMs' versatile conversational capabilities~\cite{EmbeddingLLMCUI24,ZhangVirtual24}. Kim et al.~\cite{VAsGroupChi24} investigated the impact of a screen-bound virtual agent during group tasks and found that programming more engaging behaviors like directed gaze and speaking gestures into collaborative GenAI agents captured users' attention and enhanced group synergy.

%% file: 3_Provotype.tex
\section{Research Approach and Methods}

To better understand the opportunities and challenges of embedding GenAI agents as independent contributors within synchronous group settings, we conducted an exploratory design study with 25 participants across 6 professional teams, where each team engaged in speculative workshop and an individual follow-up interviews. We chose to conduct an exploratory study that was speculative in nature instead of a controlled experiment because it encourages us to ask "what if" questions ~\cite{dunne2024speculative} and offers a powerful method for exploring the social, ethical, and practical implications of an emerging future~\cite{bleecker2022design,boer2012provotypes}. This allowed teams to engage in thoughtful debate and reflect on the possible shifts and design opportunities that would impact them as a team.

\subsection{Provotype Design}

While it was important to us to incorporate the voices of end-users in shaping a collectively desired future ~\cite{rheingold1985tools,mitrovic2021beyond}, predicting the implications of emerging technologies can be challenging for non-experts as they might struggle to envision the various possibilities, leading to vague and overly utopian or dystopian views~\cite{shorter2022materialising, morris2014reducing}. Additionally, computational intelligence is often invisible and can require more tangible experiences to draw out its potential. These challenges led us to use \textit{provo}types or \textit{provocative prototypes} in our workshops. Unlike traditional prototypes that imply a journey towards production and deployment
~\cite{to2022interactive} with the goal of developing useful features, provotypes strive to “feel out the boundaries of a problem”~\cite{carey2009tangible}. This allows for a more concrete and experiential way to provide users with a glimpse into one possible future~\cite{mogensen1992towards,boer2012provotypes}, inspiring them to consider what they want and don't want for their future selves and generate discourse around complex socio-technical systems~\cite{bardzell2012critical,celi2023prototypes}.

We developed a functional multi-user MR application that enables 3-6 participants in the same physical space to engage in a group discussion where LLM-based collaborative GenAI agents could actively participate and offer independent contributions to the discussion in real-time. Our primary design goal was to emulate a futuristic scenario with fluid conversations between human participants and GenAI agents that could spark debate. The provotype was designed to be portable and configurable to allow us to adapt to different team spaces. The collaborative MR space also included a non-interactive shared 3D artifact and a virtual whiteboard to make any space feel more like a meeting room. Our implementation details are outlined in Appendix \ref{appendix:implementation}. Below we describe our provotype's features and design rationale: 

\begin{figure} [t]
  \centering
  \includegraphics[width=\linewidth]{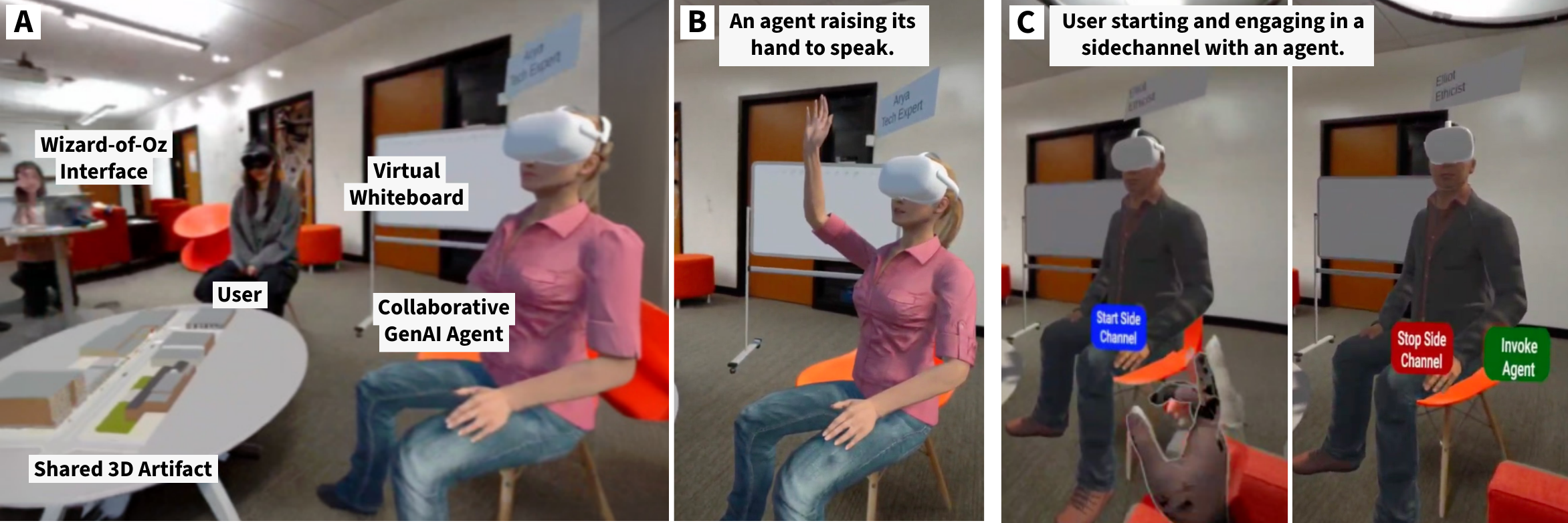}
  \caption{Our MR provotype included a shared 3D artifact of a city as well as a non-interactive virtual whiteboard. Agent contributions were generated through a GPT-based pipeline and their social behaviors were controlled through a Wizard-of-Oz interface (A). Agents could proactively indicate a desire to contribute by raising it's hand (B), and users could use MR buttons to engage in a private sidechannel with one agent (C).}
  \Description{}
  \label{fig:workshop}
\end{figure}

\subsubsection*{Discussion Topic and Agent Perspectives: } We developed and optimized GenAI agents to contribute to a discussion on participants' opinions about whether the city they lived in should deploy smart street lights with cameras and sensors. This topic was chosen because it was one that any resident could have an opinion on and a personal stake in, much like regular team discussions. Furthermore, choosing a domain-agnostic topic allowed teams to participate without risk of disclosing proprietary information they did not wish to share. We also designed the GenAI agents to embody distinct perspectives, with each contributing unique viewpoints to the group discourse. Each agent assumed the persona of a technology, sustainability, or ethics expert. 

\subsubsection*{Representation of GenAI Agents in MR Space: }
We chose to have the collaborative GenAI agents embody human-like avatars and contribute to the discussion using LLM-generated speech. A humanoid form was chosen over more abstract representations to highlight their role as an active collaborator as well as create a "desirable discomfort"~\cite{auger2013speculative} and inspire participants to think more deeply about how they would like GenAI collaborators represented in the future. However, to minimize any eerie feelings stemming from uncanny facial features, we opted to have GenAI agents wear MR headsets. This also allowed users to view the agents in a similar manner to other human discussants who also wore MR headsets. 

\subsubsection*{Number and Spatial Location of Agents: } The number of agents that participants interacted with was the equal to their group size. This allowed each participant to have an assigned agent in the individual mode (outlined below), enabling them to discuss their perspectives regarding the integration of multiple distinct agents within a team. While the GenAI agents could be configured to occupy any location within the collaborative space, our workshops participants sat around a table of their choice and agents were placed around the same table. A set up mode within the MR application allowed for researchers to configure the number and location of agents for each user.

\subsubsection*{Interacting with Collaborative GenAI Agents: }
Each GenAI agent was assigned a name to allow them to be referred to in conversation, and their perspectives were made visible to participants through a label that floated above their head so participants didn't need to remember it. These collaborative agents were also programmed with animations to make them feel more lively -- a breathing animation provided subtle movements when an agent was idle and a hand-raise animation allowed agents to indicate when they wanted to contribute to the discussion without unnecessarily interrupting it. Agents could also turn to users when they were speaking and their animations included synchronized mouth and arm movements. While the contents of the agents' speech itself was a result of a GPT-driven pipeline, a study facilitator used a Wizard-of-Oz system to control when a GenAI agent spoke or raised its hand during the discussion. This allowed for a more natural simulation of agents capable of understanding socially appropriate moments to interject and offer their generative opinions.

\subsubsection*{Individual and Collective Interaction Modes: }
To better reflect the possibilities associated with future technologies--in particular, MR's ability to create asymmetric and personalized experiences for each user within a collaborative space~\cite{gronbaek2024blended,johnson2023unmapped,nebeling2020xrdirector} -- we designed two separate interaction modes that our participant teams could try: a \textit{collective mode} and an \textit{individual mode}. In the collective mode, all the discussants could interact with all the GenAI agents purely through speech and non-verbal communications. A study facilitator observing the conversation acted as the Wizard, controlling when an agent spoke or raised their hands. 

In the individual mode, each participant could only engage with one agent that was pre-assigned to them. When seated, discussants could turn to their assigned agent and engage in a sidebar conversation; during our workshops, this was the agent to their right. While multiple participants could interact with their agents simultaneously, they could only hear the responses of their own agent. However, all participants could visually observe speaking animations from any agent. This design emulated sidechannels or breakout sessions during colocated meetings. The individual mode did not involve a Wizard of Oz component, instead giving participants direct control over when their agent spoke. To enable this, the MR interface integrated buttons to `start sidechannel', `invoke agent', and `end sidechannel'. When involved in a sidechannel, a participant's GenAI agent faced the assigned participant.

%% file: 4_Method.tex
\begin{table}[b]
\centering
\includegraphics[width=\linewidth]{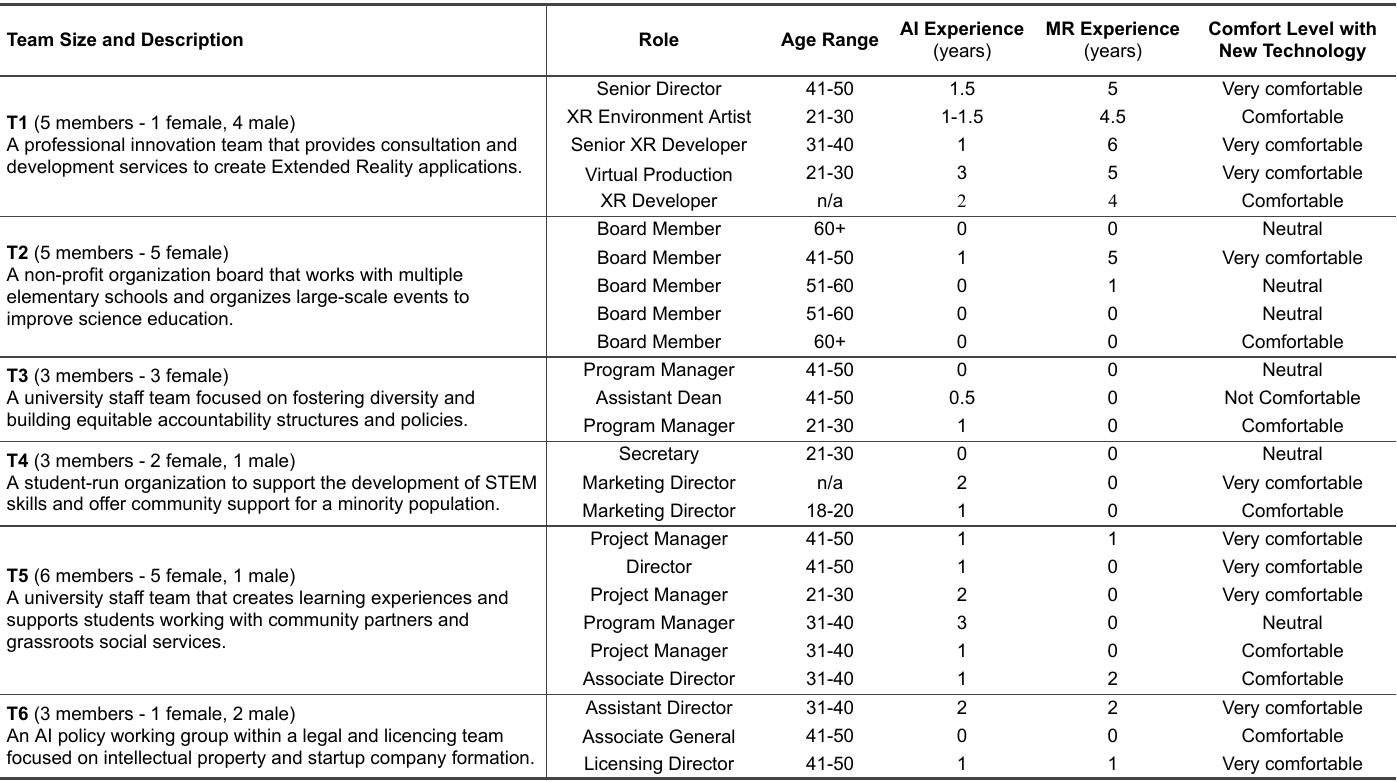}
\caption{Team and participant details for our exploratory study. Teams were chosen based on their potential to offer diverse perspectives on collaborative GenAI agents across a wide range of roles and domains.}
\label{tab:participants}
\end{table}

\subsection{Participant Recruitment}
We recruited local professional teams that tackled complex and multi-faceted problems and regularly engaged in creative ideation, problem-solving, and deliberation through synchronous meetings. Recruitment took place through a North American university and through authors' professional networks. The research team facilitated 15-minute screening calls with one spokesperson from interested teams to better understand their team functions and set expectations for the study. Teams that were deemed a good fit were then scheduled for the workshop. Researchers prioritized multi-disciplinary teams in a variety of domains to promote diverse perspectives in the study. 

A total of 25 participants across 6 professional teams with 3-6 people each were recruited (see Table~\ref{tab:participants}), leading to a total of 6 workshops. Team workshops lasted 2 hours and occurred in teams' own office spaces or in the researchers' lab space, depending on their preference. The follow-up interviews were conducted over Zoom\footnote{https://www.zoom.com/}. Of the 25 participants across 6 teams, 19 participants completed the follow up interview. 6 participants were not able to participate due to scheduling conflicts. The study was approved by the local ethics board and all participants provided informed consent. Participants were provided with compensation the equivalent of \$20.

\subsection{Procedure: Speculative Team Workshop}
First, participants completed consent forms and an initial survey regarding demographic info and technology familiarity. Then, teams were introduced to the goals of the workshop: creating an ideal intelligent collaborative space, where GenAI agents could make independent, proactive contributions toward achieving the team’s collective objectives. The two hour workshop session was divided into the following three sub-sessions with 5 minute breaks in between. 

\subsubsection*{Understanding team needs, and priming: } 
The first sub-session started with a discussion around the team's approach to creative problem-solving through synchronous group meetings and their use of technology to facilitate this collaboration. This allowed researchers to better situate team members' opinions and feedback during the rest of the workshop.

Teams were then introduced to the capabilities of GenAI and MR. To inspire participants to envision futuristic possibilities rather than being confined to their current experiences with GenAI tools like ChatGPT, we also exposed them to three concept videos. This included a clip from a movie that depicted of a popular pop cultural reference of an AI agent capable of contributing to creative problem-solving (we used a clip of Tony Stark interacting with JARVIS from the Marvel series), and two technology vision videos published by technology companies that depicted something similar with one showcasing screen-based interactions and another using immersive smart glasses (listed in Appendix~\ref{appendix:videos}). The aim was to ensure that participants had a shared understanding of the technology and to show that while the focus of their ideal collaborative space was primarily about GenAI agents, they didn't necessarily need to include holographic or MR components.

Participants then worked individually to brainstorm and sketch their ideal futuristic collaborative space (Fig.~\ref{fig:workshop}-A) using a worksheet that guided them through it and then engaged in a discussion about their ideas as a group. Examples of participant sketches are in Fig.~\ref{fig:whatif}, and the worksheet template is in Appendix~\ref{appendix:template}.

\subsubsection*{Experiencing a team discussion with the provotype: }
\begin{figure} [t]
  \centering
  \includegraphics[width=\linewidth]{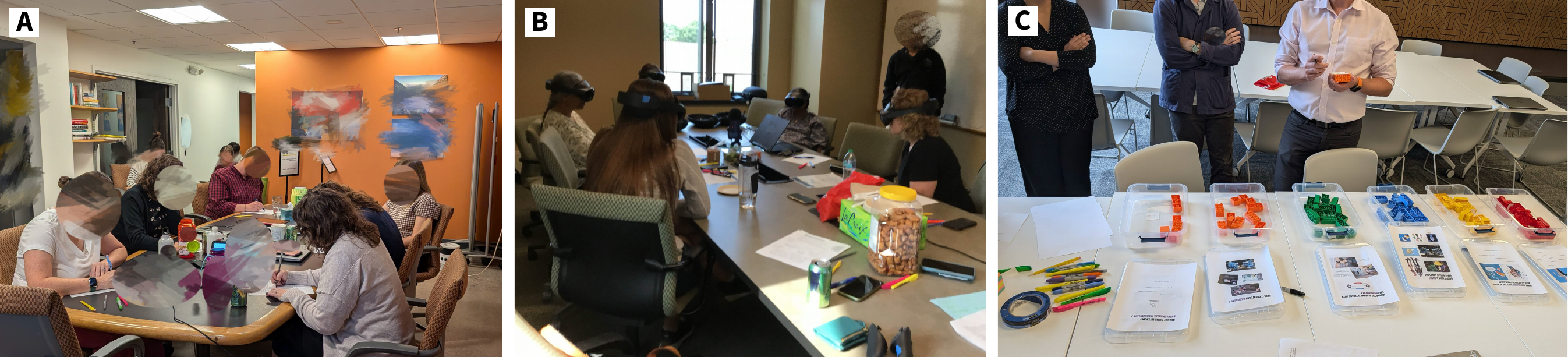}
  \caption{Images from our speculative team workshops where teams work on their individual worksheets (A), experience a group discussion with collaborative GenAI agents in Mixed Reality (B), and participate in our Lego Assembly Line activity to design their ideal collaborative agents (C).}
  \Description{}
  \label{fig:workshop}
\end{figure}

Our provotype demonstrated how GenAI agents could function as active collaborators, providing participants with an embodied experience of having agents sit around the table and actively contribute to a group discussion (Fig.~\ref{fig:workshop}-B). Participants were aided by the research team to put on the headsets and calibrate it to their eye. Each team engaged in a 5-minute discussion in the collective mode and then switched to the individual mode and interacted with their assigned agents privately for another 5 minutes before returning to a group discussion. This structure ensured balanced exposure to both interaction modes and reduced the variability in participant experiences across teams. On completing the discussion, teams were asked to provide feedback on it before engaging in a more general discussion about their perceptions of collaborative GenAI agents as active contributors and the values and concerns they saw within a future where that was a reality.

\subsubsection*{Participatory Design of Collaborative GenAI Agents: }
\begin{figure} [t]
  \centering
  \includegraphics[width=\linewidth]{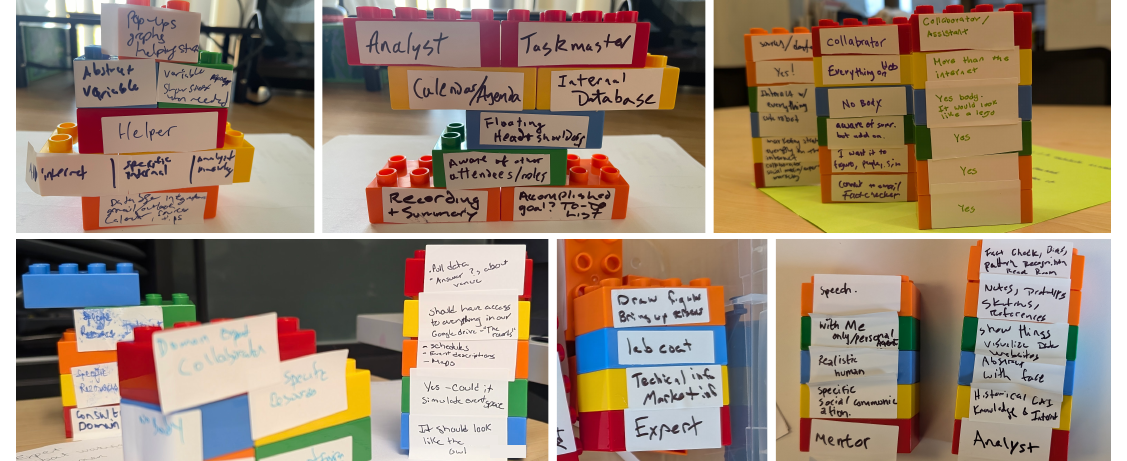}
  \caption{Example Lego agents our participants built to depict their ideal collaborative GenAI agent during our team workshops. Different colored Lego blocks were used to represent a variety of design dimensions.}
  \Description{}
  \label{fig:lego}
\end{figure}

After a short break that allowed for their experience with the provotype to sink in, teams members spent around 40 minutes on an activity called the ``Lego Assembly Line'' (Fig.~\ref{fig:workshop}-C) where they explored different design dimensions to build their ideal collaborative GenAI agent (Fig.~\ref{fig:lego}). These dimensions represented key factors that influence user experiences with computational agents and were informed both by our early explorations and prior research~\cite{bittner2019bot,holz2011mira}. This included making decisions about the agents role and knowledge, how it is represented in the collaborative space, the interactions they envisioned having with it, any interactions it was expected to have with the environment, and any additional artifacts they found useful. Team members were first asked to create their ideal agents individually before coming together and choosing which ones the team would like to use as a group. They were also encouraged to role-play specific scenarios with their envisioned agents during this discussion.

\subsubsection*{Reflecting on values within a preferred future: }
As a closing activity to the workshop, participants individually reflected on the inclusion of the collaborative agents they had come up with within their teams through a worksheet that allowed them to list their hopes and concerns for the scenarios they had envisioned. This worksheet also guided them to write a letter to makers of the future to capture the broader values they wanted to see reflected in their future. Examples are in Fig.~\ref{fig:hopesconcerns}, and the worksheet template is in Appendix~\ref{appendix:template}.

\subsection{Procedure: Individual Follow-Up Interviews}
Teams were invited to sign up for a follow up interviews 3-7 days after the team workshop. The gap allowed them to reflect on the discussions and what they experienced in the workshop as they went about their regular workdays and think more critically about it. It also allowed for researchers to dive deeper into their individual perspectives and opinions from the two individual activities and other team activities so that they could engage in clarifications and follow-up questions during the interview. The individual format allowed for team members to express their opinions without being concerned about the larger group. These interviews were conducted in a semi-structured format and lasted 30-45 minutes each. 

\subsection{Data Collection and Analysis}
The team workshops were audio and video recorded using a GoPro and a mobile phone, and participants' provotype experience was recorded from their POV on each headset. In addition to this, the research team also took pictures of the different artifacts produced from the workshop activities (the two individual worksheets and final Lego agents). The follow-up interviews were audio and video recorded through Zoom. All recordings and images were transcribed and then analyzed using Dovetail\footnote{\url{https://dovetail.com/}}. Two researchers initially familiarized themselves with the data and created an initial set of codes. They then systematically worked through data from three workshops and two interviews to iteratively and collaboratively refine the codes and create a final coding scheme. Four researchers then coded the rest of the data before discussing their insights as a team and resolving any disagreements. Finally, higher-level categories and themes were iteratively derived to converge on the results described in this paper. 

While participants frequently brought up various types of \textit{individual} AI support during the workshop, we focus our results on aspects that were relevant to having independent and autonomous contributions from a \textit{collaborative} GenAI entity in a synchronous group setting. Given the exploratory nature of this study, we also did not assess feedback on the features or the usability of our provotype, nor did we evaluate the specific interface designs suggested by participants. Instead, we examined these contributions and rationales to reveal teams' underlying needs and concerns and provide broader domain-agnostic insights.

\subsection{Overview of Findings}

The following sections present the findings in three parts grouped per our research questions: 
\begin{itemize}
    \item Section~\ref{sec:findings-rq1} examines the perceived value of GenAI agents in collaborative settings. 
    
    \item Sections~\ref{sec:findings-rq2a} and ~\ref{sec:findings-rq2b} uncover the anticipated individual and team considerations, respectively, for designing effective collaborative GenAI agents. Section~\ref{sec:findings-rq2c} then discusses broader organizational and societal considerations.
    
    \item Section~\ref{sec:findings-rq3} outlines participants' envisioned interactions with agents in MR and presents three design tensions based on their rationales.
\end{itemize} 

%% file: 5_Findings1.tex
\section{RQ1: Perceived Value of Collaborative GenAI Agents in Synchronous Group Settings}
\label{sec:findings-rq1}

Our discussions with participants during the workshops and follow-up interviews uncovered the anticipated benefits collaborative GenAI agents could bring teams during synchronous group sessions like brainstorming meetings or deliberative discussions. All our participants emphasized that collaborative agents should augment, not replace, the existing skills and perspectives on their team. In this section we outline three key opportunities teams saw for GenAI agents to enhance team outcomes and processes. 

\subsubsection*{\textbf{The opportunity to include diverse perspectives and boost confidence in team outcomes}: }

Teams saw potential in GenAI agents to enhance brainstorming and deliberation, believing they could uncover novel ideas, offer fresh perspectives, and act as devil's advocates to identify blind spots. For example, T5 noted they \textit{"often talk about one type of accessibility or inclusion [and would] value somebody who can speak deeply on the ethics of something could bring in a valuable perspective that we need"}[P22]. In a similar vein, one participant mentioned that while their team ensures \textit{"ethics is discussed as like a central tenet of every kind of decision we're making [but] I come from an industry where that is absent 99\% of the time. So I would 100\% advocate for the ethicist agents to like infiltrate every business and be forced to get 10 minutes of talking time at every meeting...I think [it] would be really exciting”}[P19].

Some teams also felt GenAI agents could \textit{"be really powerful"}[P5] for addressing social and resource disparities more effectively. T2, in particular, works with \textit{"schools that are extremely affluent and there are schools that are not"}[P7] and were excited about agents' ability to offer contributions like \textit{"hey, you might need these tools in order to help [some schools] become equal to these other schools"}[P8]. While participants acknowledged that \textit{"it would be much better to have [a real person] at the table than an AI representing somebody who might be in that situation}"[P5], teams saw GenAI agents as especially valuable in early ideation when access to real voices were limited. In general, teams saw GenAI agents as a safety net that could \textit{"give us more confidence that there isn't something we're overlooking"}[P10].

\subsubsection*{\textbf{The opportunity to challenge groupthink and build on organizational knowledge}: }

Teams recognized that working together over time can lead to ingrained patterns and limit creativity because \textit{"we all have very similar knowledge of how we’ve done things in the past [and] certain opinions we might have on topics are formed by like, how we operated"}[P1]. They envisioned GenAI to be a valuable tool that could \textit{"challenge some of the idiosyncrasies like of, well, I would say like groupthink"}[P22], helping teams broaden their outlook and break out of their established patterns. For example, P19 acknowledged they have a tendency to \textit{"just focus on how things are done in [one region]...if an agent could bring in a perspective and be like, oh in [a different region] they do this, there could definitely be elevation there"}. 

Additionally, some participants highlighted challenges with accessing past solutions due to team turnover; they were excited about leveraging organizational memory through GenAI agents that had \textit{"a storehouse of ideas [to] lean into"}[P10]. This would allow teams to \textit{"deepen [the knowledge] we already have [and] fold in new things with a balance of history...and resist this idea that it has to be new and faster and better and more"}[P16]. 

\subsubsection*{\textbf{The opportunity to bridge communication gaps and reduce social friction}: }

Teams highlighted that \textit{"it'd be interesting to have [an agent] who is not necessarily an expert in any particular type of knowledge but can spot patterns in people's conversations"}[P5] to bridge communication gaps in real-time. For example, P24 envisioned an agent who could flag excessive jargon and say \textit{"Oh, you! You made a leap there like you left out a piece of knowledge that somebody may not have"}. Some also imagined agents could offer additional resources to the group to strengthen their arguments. 

\begin{figure} [h]
  \centering
  \includegraphics[width=\linewidth]{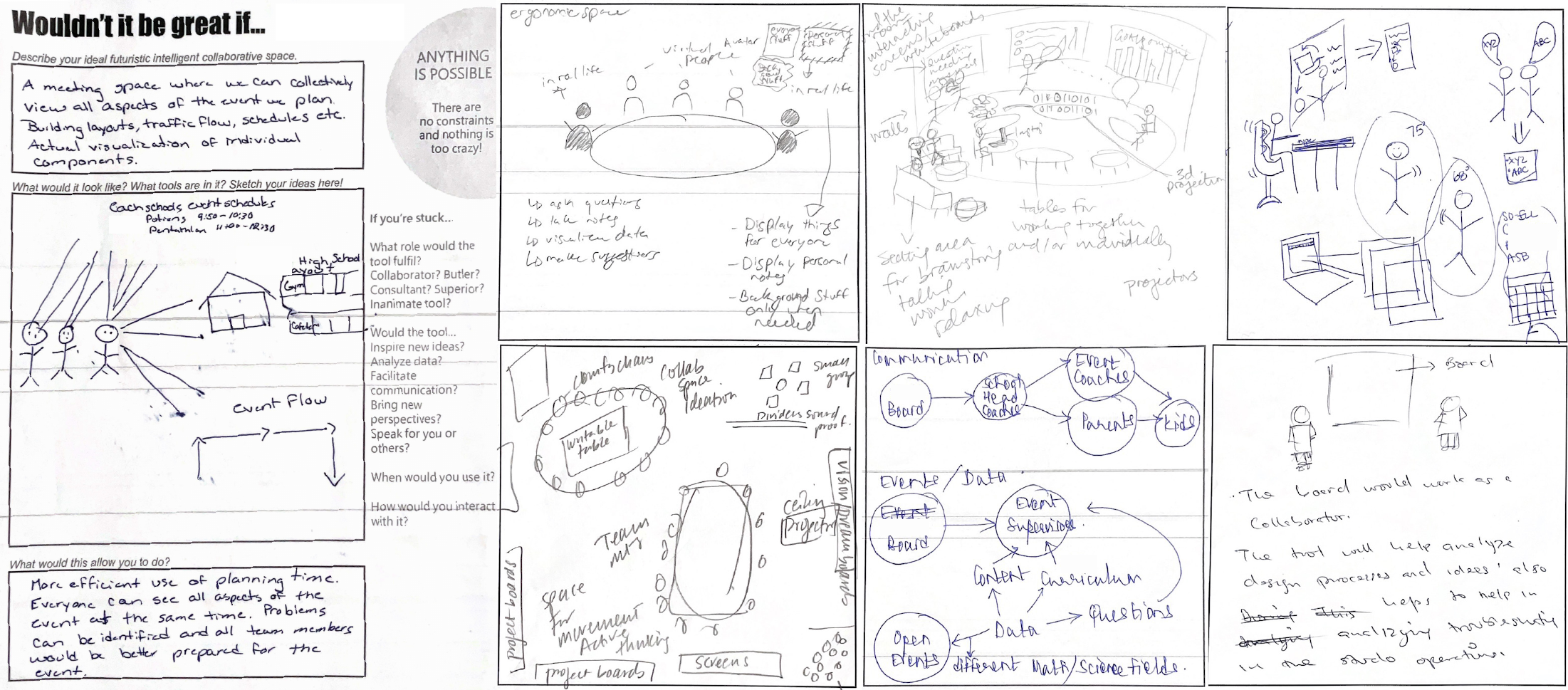}
  \caption{Example sketches of participants' envisioned ideal futuristic collaborative spaces from our workshops.}
  \vspace{-1em}
  \Description{}
  \label{fig:whatif}
\end{figure}

An agent that interprets facial expressions and body language could also serve as a neutral moderator and reduce social friction by taking on tasks that might create awkwardness or discomfort in human interactions. This included prompting quieter participants to speak, reminding vocal ones to be mindful, and mediating difficult conversations. Some participants felt this could \textit{"actually make life a little bit easier working with people if there was some kind of greasing the wheels in between all of us"}[P17]. Teams also acknowledged leaving meetings unclear on key points only to \textit{"immediately go into the hallway and ask their friend ‘so what did we just talk about?’"}[P2]. They imagined an agent could \textit{"be the awkward person in the room"}[P3] and ask clarifying questions without placing the burden on individuals to admit confusion while maintaining common ground.

A few participants also mentioned that hierarchical structures or social dynamics could lead to them being scared or hesitant to speak up and address overlooked perspectives because \textit{"humans get nervous or we're not sure that this is our role to do that"}[P21]. They imagined feeling more comfortable raising concerns indirectly by asking an agent questions that would naturally bring relevant issues to light, and felt that the perception of GenAI being all-knowing would be \textit{"helpful because the machines, the agents, are calling it out"}[P13] and they could address critical issues without worrying about exposing gaps in others’ knowledge. 

%% file: 6_Findings2.tex
\section{RQ2: Anticipated Design Considerations for Integrating GenAI in the Workplace}
\label{sec:findings-rq2}

While participants recognized the potential of collaborative GenAI agents to enhance collective capabilities, they also identified numerous risks and concerns about their future use. Much of our workshop discussions centered on these concerns and how teams envisioned addressing them. In this section, we outline three user needs each across individual, team, and broader organizational considerations. We also discuss the different underlying concerns and potential workarounds teams proposed for each. In general, participants emphasized that careful considerations across these levels was crucial to collectively shaping a team's willingness to adopt a GenAI agent. They predicted that each individual would personally assess the value of an agent for themselves before either supporting its integration into the team or discouraging its use. 

\subsection{Individual Considerations}
\label{sec:findings-rq2a}
\subsubsection*{\textbf{[I1] The need to trust GenAI agents and evaluate perceived usefulness}: }

Participants stressed that agents' contributions must be relevant, accurate, and aligned with the team’s goals to be perceived as useful and not something that hinders progress or creates confusion. A few participants likened agents to interns or new members and predicted that they would initially forgive errors but saw building trust as an evolutionary process that hinged on consistent and reliable interactions across multiple sessions of use.

Bias in GenAI models was a major concern and teams wanted transparency and assurance that the agents' knowledge came from reliable sources. P9 mentioned this was especially crucial when teams dealt with sensitive issues saying \textit{"in an age when facts don’t always matter, I’d want understand where the AI’s knowledge is coming from"} to ensure that they don't perpetuate algorithmic biases. Teams also wanted the ability to trace an agent's decision-making process. However, participants also recognized that evaluating an agent’s contributions could be challenging without expertise in its specific domain, particularly in real-time. In general, teams wanted potential biases clearly flagged to understand an agent's limitations and expected designers to incorporate indicators like labels, fact-checkers, and confidence scores within interfaces. They also expected agents to consistently remind them that it was not a definitive source of information and that its contributions reflected general or aggregate insights.

\subsubsection*{\textbf{[I2] The need for human connection and maintaining relationships}: }

Participants felt that collaborative GenAI agents could mean people rely more on GenAI contributions and feel less motivated to actively participate in collaborative settings. Few worried this could even lead to people \textit{"sending their agent instead of themselves"}[P1] to meetings, reducing the opportunities available for them to build and maintain relationships with their colleagues. A few participants were concerned that introducing GenAI agents to remote or newly formed teams where relationships haven’t fully developed could \textit{"make the people team have a weird dynamic"}[P20]. 

Additionally, while some participants felt it would be nice for GenAI agents to engage in small talk, given its prominence in synchronous meetings, they were also \textit{"torn because...I don't want to personify it too much"}[P18] because \textit{"there's a natural propensity to humanize things"}[P11]. P22 emphasized this concern when they reflected on \textit{"the concept of community [and] isolation and mental health...I think it's already clear there's a lot of harm that technology has caused, um, the human brain and human relationships, [and this could] continue to increase that right, where [people] will literally see opportunities for technology to replace human relationships and human contact"}. Ultimately, participants wanted to engage with GenAI agents without creating a false sense of intimacy and overshadowing their relationships with their colleagues.

\subsubsection*{\textbf{[I3] The need to accommodate and bridge individual differences}: }
Participants wanted GenAI agents to cater to individual communication styles, as they're already contending with a lot of information in complex work environments and didn't want to be \textit{"engulfed by these digital things"}[P11]. The relevance of the information to the individual's role was also mentioned as a factor because \textit{“in general, those of us who are not involved in a particular aspect will often just kind of check out anyway, of that aspect of the discussion"}[P9]. They wanted a \textit{"simpler system [where you have] exactly what you need for your [individual] work productivity”}[P24].

Some participants also highlighted that collaborative settings can include \textit{"a lot of distractions, especially for neurodiverse folks"}[P13] and while \textit{"some people prefer a long verbal answer, but I don't...so without bullet points, you've lost me"}[P8]. They emphasized that agents needed to accommodate individual abilities and sensory needs for its contributions to be truly impactful.

\subsection{Team Considerations}
\label{sec:findings-rq2b}

\subsubsection*{\textbf{[T1] The need for social sensitivity and fit to team culture}: }
Teams wanted collaborative agents to exhibit social awareness and sensitivity, like knowing when to interrupt and detecting tone to gauge when it's appropriate to offer input. Agents were also expected to raise contentious points \textit{"without belittling as that would be problematic [and] counterproductive to the dynamics"}[P23]. Additionally, participants wanted to ensure agents were inclusive of everyone on the team and emphasized the need for the underlying technology to be unbiased. For example, P3 mentioned \textit{"if it's unable to recognize a person of color...like the cameras don't pick that [or] if it's not trained properly...having that one person left out and the others are recognizable...that's not okay”}.

In our participants' envisioned future, GenAI agents also needed to blend seamlessly into their team’s unique culture and pick up on aspects like their humor. T5 also felt agents should mirror the team’s mood, saying, \textit{"If we're all feeling lousy, we want [the agent] to feel lousy too"}[P21]. Teams expressed that they work hard to form a culture and because \textit{"we've taken on the idiosyncrasies of this group and the way we work may be different from [others]. If an agent can't bring those in and recognize them, like immediately, I can imagine we're like, oh not invited to the next meeting...turn that off"}[P17].

\subsubsection*{\textbf{[T2] The need to protect human workflows and support critical thinking}: }

While teams valued agents that could proactively offer input, they worried that GenAI contributions could be disruptive to their creative process and wanted the option to toggle it on or off, or for agents to signal when they have input but remain on standby until prompted. Participants were also concerned about becoming overly reliant on GenAI and felt that integrating readily available collaborative agents would make it very easy for them to turn to it without sufficient considerations because \textit{"if the AI thing just has the right answer immediately...like, why do we need to brainstorm?"}[P25]

Participants also emphasized wanting to protect the human element of \textit{"being able to bounce ideas off of each other...I just love that collaboration when it comes to building solutions or strategy”}[P11]. Engaging in the process was also seen as crucial for skill-building and especially necessary for junior members to develop their own problem-solving abilities. Reflecting on these concerns, P21 noted that \textit{"I want a tool that's clever, but I also enjoy finding the answer and feeling clever"} and envisioned being able to place \textit{"some constraints [on the agent] to say, um you know, don't give us the answer, you can only ever get us X percent of the way...get [us] 50\% of the way there, get [us] 80\% of the way there"}.

\subsubsection*{\textbf{[T3] The need to control decision-making and manage AI influence on team outcomes}: }
Teams felt strongly about keeping decision-making firmly in human hands and limiting an agent's influence on their outcomes. Participants' current experience with GenAI capabilities impacted this view -- \textit{"I mean, just the way some of it is now it's so confidently wrong, right?...I wouldn't want it [making decisions]"}[P1]. Reflecting on the multifaceted and social aspects they work with, P21 felt that \textit{“humans would be able to [act with] grace and leeway [but] a computer is not going to do that, right? It has a goal and a set of parameters it's gonna pursue those...I think it could get to a point where it pushes people to act [not] in the benefit of the team, but rather in the benefit of whatever goal the AI has been given...from its perspective, the people are also just tools, a means to an end...I can see that going south in bad ways”}. 

Participants emphasized that while the AI agent should be available when needed, it shouldn’t have equal prominence to human team members and envisioned GenAI agents to take on a subordinate role and be at a lower hierarchical level than everyone on the team as one way to control its influence. Moreover, some participants needed control of the agent to remain within the team and not at the organizational level. For example, P20 mentioned organizations can have \textit{"centralization of some of the tools [that becomes] a bottleneck. Like we can't always do the things we need to because we have to go ask permission. And so, if that was a thing too where like we couldn't adjust the agent...I think we would just stop using it"}.

\subsection{Broader Organizational and Societal Considerations}
\label{sec:findings-rq2c}

\subsubsection*{\textbf{[O1] The need for additional workflows, training, and guidelines}: }

Teams expressed a willingness to be accountable for decisions involving GenAI, but highlighted that this would require additional workflows and human oversight because ultimately \textit{"it's the humans that need to figure out the right balance of perspective"}. This included adding workflows for fact-checking and research to ensure agent contributions are accurate and contextually appropriate. A few of the managers in our workshops also noted that unlike when GenAI is used as an aid for task completion, engagement with collaborative agents influencing opinions would require more deliberate reporting structures. Teams also highlighted the need to rethink recruitment strategies, noting that the widespread deployment of collaborative agents could reduce individuals' motivation to contribute. For instance, P9 mentioned \textit{"[in a volunteer-based] non-profit organization such as ours….Would people feel like, well, you've got that now, you don't need me. Um, will it reduce, um, people's willingness to step on the board because they think they're not needed as much?...I'm afraid that we might reduce our, um you know, being able to recruit new people"}.

Participants also emphasized that without proper training, \textit{"people are gonna skip to like the immediate gratification like with ChatGPT [where] they'll go on and start using it"}[P14] without fully understanding its limitations. While the ability to bring diverse perspectives to the table could be very beneficial, as P13 emphasized, \textit{"diversity is always gonna drive us to more innovative perspective building, but you only get benefits of diversity if you know how to work through the challenges of it first"}. A few participants also predicted that people would dismiss a valuable perspective simply because it came from an agent or take an agent's opinions over their teammates', saying \textit{"my agent who has access to all the latest research...told me it's fine. So who cares what you think?"}[P11]. They felt that there would need to be clear organizational guidelines and established standards about how to consider GenAI contributions that also outline respectful etiquette around agent interactions in collaborative spaces.

\subsubsection*{\textbf{[O2] The need for psychological safety within work environments}: }
Teams expressed concerns about privacy and the potential misuse of AI systems for employee surveillance, fearing it could undermine psychological safety at work. Participants mentioned the need for GenAI agents to respect their privacy and for them to feel safe to have private conversations without worry because \textit{"sometimes you're a little bit more frank about a situation [in a smaller group], than you would if the, if the external group were there...so if there were any concerns of that getting out...conversations might tend to become more formalized"}[P23]. 

Participants also mentioned needing assurance agents were not \textit{"sending [reports] out to my supervisor to, let's say, to do my review"}[P3]. On their final worksheet P21 mentioned \textit{"ultimately we all just want to do good work and go home. Help us do that, and we will love it. Create tools for monitoring tracking, and oversight, and we will hate it"}. Additionally, teams emphasized the importance of respecting individual contributions and were worried about how credit would be handled, particularly when GenAI agents play an active role in collaborative work. They feared a policy where agents were considered company property and compared it to existing policies where anything created within the organization is claimed as it intellectual property. In general, participants felt \textit{"while its fair to acquire money for the business, don't do it at the cost of total disregard to human rights"}[P12].

\subsubsection*{\textbf{[O3] The need for responsible and thoughtful deployment}: } 

Teams emphasized the need for a responsible, well-prepared approach to launching collaborative GenAI agents. For example, P13 mentioned \textit{“how do we launch something like this and make sure that I guess the world is ready for it...I feel like ChatGPT just got launched so fast and people had no idea what to do with it...[a lot] could have been curbed if there was just, like, more consciousness about launching it. So if we were to do something like this...we would need to think of all the impacts and policies before a launch instead of being reactive”}. When discussing the broader implications of integrating collaborative agents, teams also reflected on the fact that \textit{"power is a lot about who has access to information"} and expressed concern about how technology could exacerbate privilege gaps and widen the digital divide. 

Overall, teams were hopeful about the future and wanted practitioners to \textit{"be careful, be respectful, be equitable, while still being revolutionary"}[P25]. One participant wanted to use this an an opportunity to \textit{"start using these tools in these ways to show proof of concept because we know that the other folks are using it [to] maximize profit...[we can] imagine a different type of organization...I think the word empowerment keeps coming back to me but I think [we can create] an empowered organization [that uses] this in thoughtful and reflective ways that really focus on the common good"}[P12]. 

\begin{figure} [t]
  \centering
  \includegraphics[width=\linewidth]{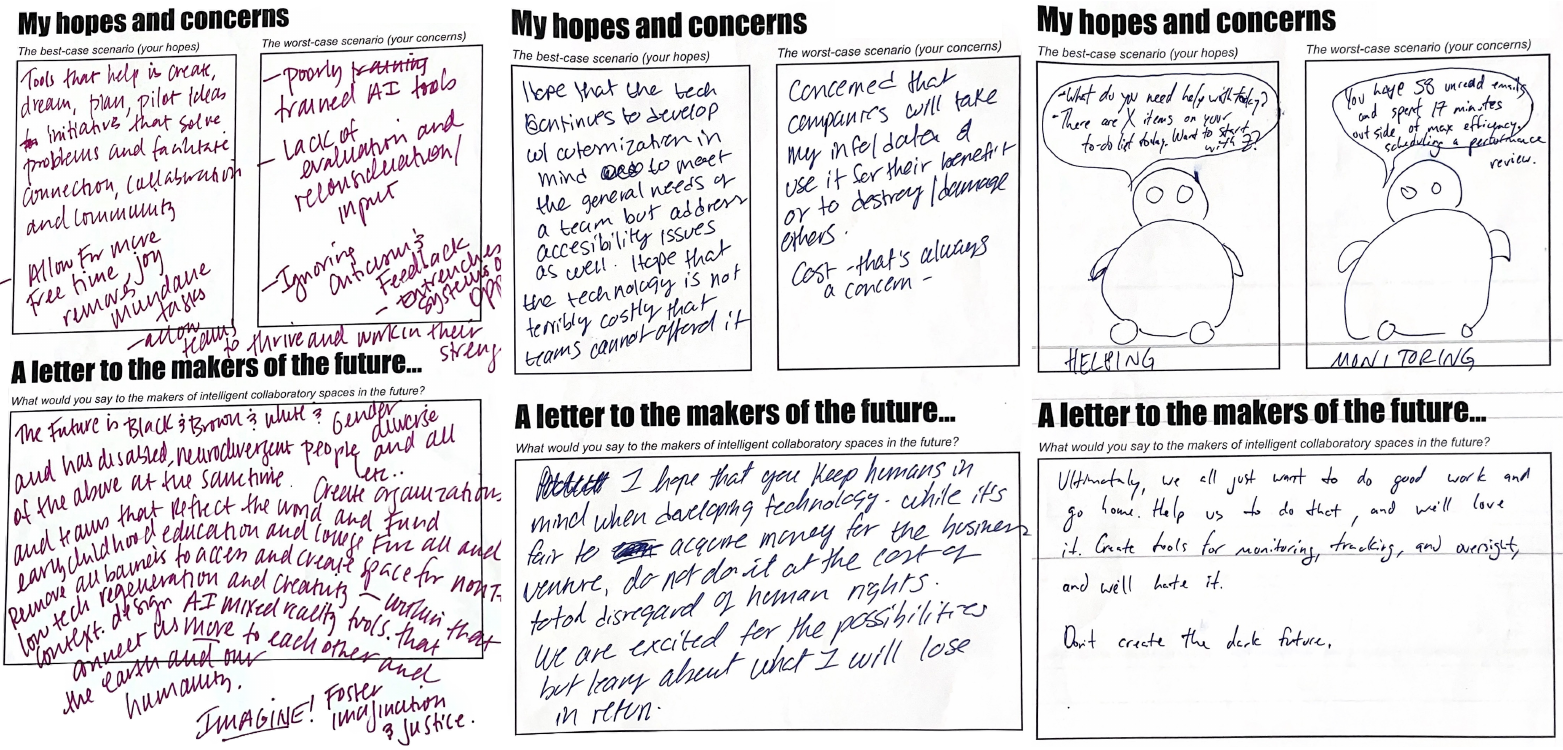}
  \caption{Example worksheets from our team workshops where participants reflected on their hopes and concerns with respect to collaborating with GenAI agents. They also wrote letters to future leaders and designers that communicate the broader values they want to see reflected in their future.}
  \Description{}
  \label{fig:hopesconcerns}
\end{figure}

%% file: 7_Findings3.tex
\section{RQ3: Anticipated Design Tensions of Interacting with Collaborative Agents in MR}
\label{sec:findings-rq3}

\begin{figure} [p!]
  \centering
  \includegraphics[width=\linewidth]{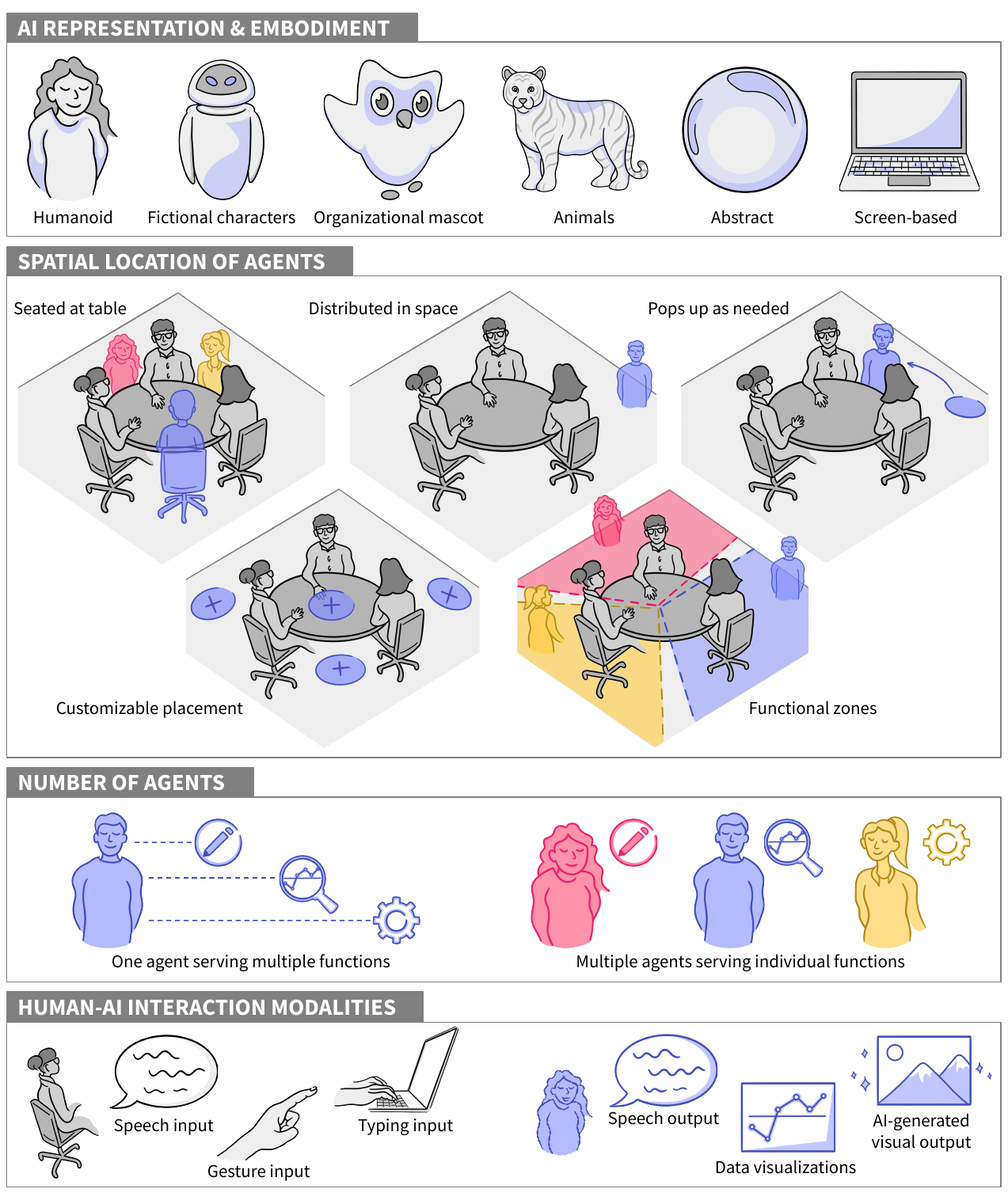}
  \caption{An artistic rendition of the different interface elements participants envisioned in our team workshops across agent embodiment, their positioning in the collaborative space, the number of agents present, and the key interaction modalities they preferred when interacting with collaborative GenAI agents. Screen-based representations and interactions were seen as helpful for individual agent interactions if more privacy was needed or to make sidechannels less disruptive to the rest of the group. Participants also wanted collaborative GenAI agents to have the ability to supplement speech output with additional information when relevant.}
  \Description{}
  \label{fig:interface}
\end{figure}

Our Lego prototyping and role-playing activity also provided insights into how participants envisioned future interfaces to address their needs outlined in the previous section. In general, most participants appreciated the idea of engaging with GenAI agents in more embodied ways during group settings like colocated meetings. The different types of interface elements teams proposed are depicted in Figure~\ref{fig:interface}. In this section, we outline three key design tensions participants' rationale for these interface elements revealed. 

\subsubsection*{\textbf{Design tension \#1: Expressive and humanoid agents could be seen as more intelligent and collaborative, but implicit biases and users' lived experiences might influence how much they trust and integrate GenAI contributions.}\\\nopunct}

During our workshops, the visual embodiment of an agent came up as a factor that had an additional layer of meaning to users' expectations of the agent's role. For example, a few teams chose organization mascots for agents that had organizational knowledge because it would be a \textit{"perfect a vehicle for someone bringing in more knowledge to assist us"}[P8]. Teams also designed fantasy characters or animal avatars to bring a sense of whimsy and fun to the workplace. However, some participants acknowledged \textit{"I definitely wouldn't take an animal as seriously...I want to take this [agent] seriously...So I feel like if they were like a cute thing, I would feel like maybe I wouldn't have the same respect as I do for another person"}[P19]. Some participants also envisioned each team member being able to personalize their agent embodiment to their preferences.

In general, most participants felt abstract and non-humanoid representations were better suited for agents acting as aids rather than collaborators because it wouldn't be perceived as having human-like intelligence. That said, participants cautioned that humanoid agents pose a risk of fostering misplaced trust because \textit{"the information becomes more compelling or more convincing when it's like a person telling you"}[P15], and the desire to be reminded that the agent is a computational entity was a major reason why some participants preferred non-humanoid characters. This transparency was seen as particularly important in settings like virtual meetings, where both humans and AIs might use similar avatars.

A few participants noted that their reaction to GenAI contributions could also depend on their own experiences and the perceived authenticity of the agent's humanoid form. For example, when P12 reflected on an agent in our provotype bringing up biases towards people of color, they envisioned dismissing the agent's contributions because \textit{"as a [POC] in that space...listening to them talk about my identities...[reminded me of others] talking about [POC] without the lived experience of it"}. However, P13 then mentioned that if the agent's avatar reflected a different cultural background to the perspective they were bringing in, it could serve as \textit{"a reminder there to not take them at face value, um, for the experience that they're sharing because they don't look [like they have lived it]"}. 

\subsubsection*{\textbf{Design tension \#2: Higher prominence of GenAI agents in a collaborative space could enhance engagement, but it might encourage overuse and lead to unnecessary influence on team outcomes.}\\\nopunct}

Participants felt that having an agent in their space was more engaging and its mere presence within a collaborative space would serve as a reminder to take its perspective into account even if the agent remained silent. They also noted that multiple agents would allow them to assign specific roles to specific spatial positions which would make it easier to intentionally engage with certain perspectives. However, participants also discussed how multiple agents could feel overwhelming. The physical prominence of an agent would also inherently put pressure on them to engage with it more frequently, saying \textit{"it would be weird just to have it, like sitting there if I'm not talking to it all the time"}[P19]. Additionally, having the agents seated at the table in the provotype discussion made some participants feel \textit{"like [they] were giving the avatars equal footing as the humans"}[P16]. When reflecting on this, participants mentioned \textit{"I don't think [agents] necessarily had to be at the right round table themselves...I don't want it to be as prominent as people, but having [it] present in the space but not equal in the space [is] probably the safe spot for me”}[P21].  

To counter this, a few participants envisioned the agent would only pop up when needed. However, most felt this would remove the ability for the agent to be proactive and imagined being able to configure the location of the agent and move them farther away in the room. One participant imagined collaborative agents being given a dedicated, persistent space within an office setting, saying \textit{"I like the idea of like the AI being, maybe there's a separate room or maybe it's in the middle of the room...Um you know, so like if anybody has any questions [or] maybe just like an idea you have to get up from your desk, go over, talk to it”}[P15]. They felt that this would allow for a signal for others to know how much they interact with the agent, and add social pressure to avoid appearing overly reliant on the agent.


\needspace{3\baselineskip}
\subsubsection*{\textbf{Design tension \#3: Shared agents could allow for easier detection of GenAI errors, but interacting with individual agents privately or in sub-groups could allow humans to filter GenAI contributions and manage agent influence on team}.\\\nopunct}

Participants generally felt having shared collaborative agents in a group setting (collective mode) facilitated natural conversation and provided equal access to information. P1 noted that they often look up information during meetings to benefit the group but end up miss ongoing discussions and having shared agents would \textit{"allow it to come out in conversation, which is a much more natural way for everyone to hear”}. Teams felt a major advantage of shared agents was easier detection of errors in GenAI contributions, as it only required one skeptical member to prompt the entire group to think critically -- \textit{“I prefer where we could all connect with our agents because we could also speak to each other and like bounce off ideas or question uh whether we agree or disagree with that information that was being given to us and we can process the response together...[it decreases] the opportunity to just believe blindly"}[P11].

On the other hand, participants imagined the ability to interact with a collaborative agent individually would allow them to process information and refine their thoughts privately before sharing them with the group in real time. P20 mentioned \textit{"I think there is value in that because [it's a] very natural human interaction that happens in a group of more than three people is there will sometimes be a sidebar conversation about something...sometimes it's, you wanna like rapidly go through this idea you have before you bring it to the larger group}". P3 echoed this saying \textit{"it sort of like, mimics the Zoom breakout groups. So it could be like, you know, you are like a larger group of a 10, a group of 10 people and then you break out into a smaller groups and each has [assigned agents]...the value is to be able to [say] `OK, you're gonna talk to this expert and get this information, and then we're gonna come back to the table and do this'"}. Whether through individual conversations or breakout groups with collaborative agents, participants saw different interaction modes with GenAI as a way to manage agent influence on the team. For instance, P19 mentioned interacting with agents privately or in smaller group would be preferred in some scenarios\textit{"because then it's not, you know, influencing the whole group...So it's not swaying like the whole group because I like having diverse perspectives"}.

%% file: 8_Takeaways.tex
\section{Main Takeaways and Limitations}

Engaging diverse professional teams through a provotype that offered an embodied experience of a potential future allowed us to engage in critical discussions and debate about collaborative GenAI agents with end-users. In this section, we summarize and reflect on our findings with respect to our research questions in Section 1. 

It is worth noting the potential sampling bias introduced by the novelty of the technologies explored. For instance, while we did not explicitly recruit participants based on their openness to new technologies, the focus on emerging technology likely resulted in self-selection bias by attracting individuals inherently more receptive to it. As a result, our study may not fully capture the perspectives and concerns of those more resistant to adopting new technologies. Additionally, our study design centered around understanding end-users and ensuring limited technology experience did not hinder participation - but despite our best efforts, the relatively low prior experience most participants had with AI and MR meant they may have overlooked aspects experts and more experienced users would have considered. Moreover, the insights discussed in Section ~\ref{sec:findings-rq3} were shaped by participants' exposure to our MR provotype, and therefore capture only a subset of the possible design tensions with respect to the design of collaborative GenAI technologies.

\needspace{2\baselineskip}
\subsubsection*{\textbf{[RQ1] Perceived value to enhance collective capabilities (Section~\ref{sec:findings-rq1}): }}
Participants felt collaborative GenAI agents could enhance problem-solving by offering diverse perspectives, act as a safety net by identifying blind spots, and reduce social friction by bridging communication gaps. This aligns with prior research where GenAI was recognized as a promising way to augment perspectives~\cite{he2024ai, chiang2024enhancing, de2023ai, inie2023designing}, manage group dynamics ~\cite{rayan2024exploring, he2024ai}, and build common ground~\cite{suh2021ai}. Our findings also highlight the potential for collaborative GenAI agents to help teams navigate power dynamics by acting as a neutral intermediary or a "scapegoat" that awkward interactions can be offloaded to. In general, teams felt that, if designed right, collaborative GenAI agents could empower teams to overcome teamwork limitations like groupthink and silos while balancing innovation with broader perspectives and historical knowledge. 

\subsubsection*{\textbf{[RQ2] Acceptance of Collaborative GenAI agents in  group settings (Sections~\ref{sec:findings-rq2}): }} Our participants emphasized that a team's adoption of collaborative GenAI depended on its fit at both individual and team levels, with individual perceptions of trust and usefulness~\cite{shen2024towards} remaining crucial in group settings as well. They also expected to assess an agent’s potential impact on human relationships and ability to bridge differences before endorsing its integration into the group. 

On a team level, our findings reinforce the need for GenAI agents to demonstrate socio-emotional competence in group settings~\cite{siemon2022elaborating, chiang2024enhancing}, and also highlight an agent's alignment with team culture as a key factor in it's adoption. Teams also emphasized the need to collectively control decision making and protect collaborative problem-solving workflows. Finally, our participants emphasized the need for responsible and equitable deployment of collaborative GenAI, advocating for workplace psychological safety through careful privacy protections and intellectual property policies. They also emphasized that GenAI's relatively more active role in problem-solving would necessitate new workflows, reporting structures, and clear guidelines on etiquette and usage. 

These findings mirror prior work that indicates the inclusion of GenAI agents could reduce people's willingness to engage with both the process~\cite{chiang2024enhancing} and with each other~\cite{suh2021ai}, and the growing calls to protect human workflows and critical thinking when using GenAI to boost individual productivity~\cite{woodruff2024knowledge,lee2025impact,seymour2024speculating}. The need for organizational and broader structures to be in place also echoes the need for bidirectional alignment~\cite{shen2024towards} in collaborative use of GenAI as well. 

\subsubsection*{\textbf{[RQ3] Opportunities and design tensions with MR (Section~\ref{sec:findings-rq3}): }} 
Our results suggest that the spatial and immersive nature of MR allows for richer interactions with collaborative GenAI agents, complementing existing research on embodied GenAI in MR~\cite{aikawa2023introducing, bovo2024embardiment}. However, participants' envisioned ideal agent representations and interactions revealed three key design tensions: First, while the embodiment of agents could potentially enhance engagement, it also raises concerns about fostering misplaced trust, particularly with humanoid representations. Second, the prominence of agents in the collaborative space influences their perceived availability and could encourage more engagement, potentially leading to over-reliance. Lastly, the choice between shared and individual agent interaction presents a trade-off between the ability of teams to participate in the collective evaluation of agent contributions \textit{vs.}~individual agency and control over agent influence on team outcomes. While these design tensions may also extend to other technologies, we discuss their implications with a focus on MR to more completely answer RQ3 in the next section.

%% file: 9_Discussion.tex
\section{Discussion}

Our study revealed that while participants viewed collaborative GenAI agents as potentially valuable, they also identified significant risks related to both their functional capabilities and ability to navigate complex group dynamics. Effectively leveraging GenAI to enhance collective capabilities while ensuring its safe and meaningful integration into teams will therefore require the thoughtful design of both collaborative interfaces and work practices.

Several hurdles remain in realizing the vision of collaborative GenAI agents. For instance, ensuring robust model responses, real-time latency for multi-user interactions, and responsiveness to social cues are still a challenge with GenAI today~\cite{cui2024ai,ma2024towards,chiang2024enhancing}. However, the functional and social capabilities of GenAI models~\cite{park2023generative,leong2024dittos,leng2023llm} as well as its capabilities with respect to embodied interactions~\cite{bovo2024embardiment} are evolving. Progress in commercial tools like real-time voice interactions with GPT-4o\footnote{\url{https://openai.com/index/hello-gpt-4o}} and visions for independent contributions from LLM agents in meetings\footnote{\url{https://youtu.be/kkIAeMqASaY?si=vZCRiiRa3UjRQUns}} and productivity tools\footnote{\url{https://deloi.tt/3V1WdV6}} also indicate that a future with collaborative GenAI agents is not only plausible but rapidly approaching. The core challenge ahead of us is therefore a fundamentally socio-technical one.

\subsection{The Added Risk of Integrating GenAI into Synchronous Group Settings}

The cognitive~\cite{lee2025impact}, social~\cite{leong2024dittos, woodruff2024knowledge, kirk2025human}, and ethical risks~\cite{gabriel2024ethics,rauh2024gaps, weidinger2021ethical} of GenAI have become increasingly evident with its real-world use. While the inherent nature of groupwork -- such as the ability to collectively engage in critical evaluation of agent contributions -- could help mitigate some of these risks, our findings highlight that the active engagement of GenAI within teams could also introduce new challenges that need to be addressed.

Despite its promise, introducing diverse perspectives through GenAI might not always yield beneficial outcomes. GenAI models are fundamentally trained on dominant narratives~\cite{van2020post,kirk2025human} and historical data, which means they can struggle to generate viewpoints beyond their knowledge or those that require different reasoning frameworks. The particularly persuasive nature of GenAI~\cite{kirk2025human} along with group tendencies like fixation~\cite{kohn2011collaborative} on its contributions or reluctance to challenge agents~\cite{leong2024dittos} could actually exacerbate groupthink \cite{tversky1974judgment}, making it even harder for teams to break free from narrow perspectives. Additionally, these pitfalls can be particularly harmful if teams overvalue an agent's input or use it as a substitute for lived experiences as a fundamentally reductionist and biased lens~\cite{valenzuela2024artificial} could become what a population is understood through.

Our participants' concerns also highlight how collaborative agents could disrupt team dynamics. For example, fear of oversight and lack of perceived safety in the workplace could lead to rigid social norms that discourage free-flowing ideas and reinforce structures that shape whose voices are heard. Offloading difficult social interactions to collaborative agents might also weaken interpersonal relationships and eventually diminish the ability for teams to navigate conflict. Human interventions may also come to feel unusually abrupt or rude if GenAI-mediated conflict resolution becomes the norm and diminishes group cohesion. 

Moreover, while GenAI-related labor displacements are anticipated~\cite{woodruff2024knowledge, eloundou2024gpts}, our findings also suggest the presence of seemingly competent GenAI agents could discourage people from signing up for roles, especially in settings like non-profits that rely on volunteers. A decline in human expertise available for critical oversight would be particularly harmful in contexts where tacit knowledge and lived experiences are indispensable.

While speculative, these concerns highlight the risks of unchecked GenAI's influence on groups. Our findings indicate that end-users actively seek signals at multiple levels that reassure them of their safety and ability to control the influence of machine intelligence on their collaborative workflows, relationships, and outcomes. We believe the key user needs and design considerations at individual, team, and organizational levels outlined in Section~\ref{sec:findings-rq2} provide a nuanced understanding of these signals and can serve as a guide for system and process designers working on collaborative GenAI systems.

\subsection{Opportunities to Modulate GenAI Influence on Groups}

Our overall findings and the design tensions uncovered in Section~\ref{sec:findings-rq3} highlight how various design choices can influence how effectively GenAI can augment team capabilities without stifling their creativity or agency. This places a considerable responsibility on system designers to make informed decisions about the integration of GenAI agents into team workflows. However, the complexity of the socio-technical and ethical challenges involved means we do not fully understand how these decisions might shape group outcomes, team dynamics, or the broader work culture. Therefore, HCI and CSCW researchers have a vital role in advancing research that addresses these uncertainties to create a more desirable future that mitigates concerns such as those raised by our participants.

While many existing efforts to foster trust and manage AI reliance focus on \textit{explicit} interface cues and signals like confidence scores or explainability features~\cite{2023arXivTerryInteractive, shen2024towards, cao2024designing}, they may not be as effective in synchronous group settings where users need to make valid and appropriate judgments in real time. This is primarily because (1) tackling complex problems as a group is already an information-dense setting, and contending with explicit GenAI-focused cues in addition to communicating with other people and dealing with the task at hand will likely be too overwhelming; and (2) teams are likely to use GenAI to introduce perspectives not already at the table and might lack the expertise required to assess the relevance or reliability of agent contributions accurately.

Conceptual metaphors in AI design have served as powerful mechanisms that \textit{implicitly} guide users to develop mental models of computational intelligence by drawing parallels to familiar concepts like established personas and real-world roles~\cite{khadpe2020conceptual, jung2022great, desai2025toward}. Our findings suggest MR offers the opportunity to integrate implicit cues that draw on users' contextual and socially learned associations to reinforce desired metaphors or act as cognitive forcing functions~\cite{buccinca2021trust}. More specifically, with spatial and immersive technologies, these signals and affordances could leverage users' embodied and situated cognition~\cite{kirsh2013embodied} to help individuals and teams become more appropriately reliant on collaborative GenAI agents.

In the remainder of this section, we discuss \textbf{the opportunities for MR to function as a modulator through which the influence of GenAI on teams can be regulated}, and provide a future research agenda to guide both end-users and practitioners in making informed decisions about effectively integrating GenAI into teams.

\subsubsection*{\textbf{Influencing Agents' Perceived Role: }}
A collaborative agent's visual representation shapes users' first impressions of it and could significantly influence the perceived role of GenAI within the team. For example, regardless of its intended function and actual capabilities, abstract or non-humanoid forms might lead users to view an agent as a tool or assistant, while humanoid forms could mean users think of it as a peer or even a mentor. This presents an opportunity for system designers to strategically use embodiment to align the agent's perceived role with its actual capabilities. For example, representing an agent as an abstract object could reinforce its function as a tool~\cite{sarkar2023enough}, thereby modulating its influence on the team. Additionally, signaling varying degrees of cultural alignment with the team could also serve as a subtle cue that influences whether an agent is perceived as an internal or external stakeholder. 

While our participants opted for non-humanoid forms to remind themselves that GenAI agents lack true human intelligence, they also suggested that it might be necessary in certain scenarios to fully engage with and benefit from GenAI integration. However, realistic humanoid avatars have been shown to heighten trust and emotional attachment~\cite{akbulut2024all,rashik2024beyond}, which can backfire and make teams too susceptible to GenAI influence. Moreover, users' lived experiences and their resulting assumption of an agent's identity can impact how authentic different members of the group believe GenAI contributions to be, making it challenging to prevent the unintentional propagation of human bias with agents that show human-like characteristics.

Understanding how to represent GenAI agents in collaborative settings to foster appropriate reliance in groups will be a fundamentally intricate task that is heavily influenced by both group composition and the specific context of use. While prior research has examined the impact of visual embodiment~\cite{Kyrlitsias2022, GuopresentationSocialVR21, jung2022great, rashik2024beyond} and metaphors that communicate varying roles and levels of AI competency~\cite{kirk2025human, khadpe2020conceptual, jung2022great} on user perceptions, we propose that future research should explore how these effects interact with the generative capacities of AI agents in synchronous group settings, and how factors like the diversity in individual users' lived experiences can shape overall team perceptions of a GenAI agent.

\subsubsection*{\textbf{Moderating the Social Prominence of Agents: }}
Our findings suggest that an agent's social and spatial presence could significantly affect its perceived access, which refers to how available and easy to use a collaborative GenAI agent appears to be. The socio-spatial behaviors of collaborative GenAI agents in MR can therefore be intentionally designed to encourage appropriate reliance. The number and spatial positioning of GenAI agents in the collaborative environment, their non-verbal gestures, and the amount of space they occupy could all signify the prominence of collaborative agents within the group~\cite{carney2005beliefs,wessler2022virtual} and impact how integrated GenAI becomes in team activities and discussions.

While increased spatial and social presence is often seen as beneficial in immersive systems~\cite{billinghurst2002collaborative,johnson2023unmapped}, our findings suggest that this may not always be true for collaborative GenAI agents. For example, a less prominent agent may be preferable to subtly encourage users to problem-solve with each other before turning to GenAI. Adjusting an agent's spatial prominence in the environment -- by placing it on the periphery of teams' proxemic zones, or in a dedicated area that requires users to physically move towards it -- could introduce an interaction and social cost that counteracts overuse by adding a layer of intentionality and accountability to engaging with the agent. On the other hand, insufficient prominence could also hinder usability and distort GenAI's perceived value within the group. 

The spatial arrangements of people in an environment fundamentally shape the nature of social and group interactions within it~\cite{kendon1990conducting, edward1966hall} and can be designed to foster distinct collaboration patterns~\cite{wong2024practice}. Designers could also leverage this to adjust an agent's spatial prominence in the collaborative space. For example, while a roundtable setting would suggest everyone has an equal voice, a fishbowl arrangement with agents on the outer circle might signal to users that human voices should remain the primary drivers of a discussion. Further research is needed to explore how the nuances of different spatial arrangements impact the influence agents have on individual users and teams. This can then inform the design of systems that strike a balance between introducing a cost to accessing the agent and ensuring it remains readily accessible without coercing users to engage with it.

\subsubsection*{\textbf{Allowing for Collective Boundary Regulation:}}
System designers should also equip teams with tools to tailor an AI agent's influence to fit team needs and allow users to set limits and create boundaries~\cite{altman1975environment} that regulate how much influence collaborative GenAI agents have on group outcomes and creative problem-solving processes. For example, options to adjust how much an agent can intervene would preserve a teams' autonomy and allow GenAI agents to assist without overstepping or overwhelming them. Another approach that leverages the flexible collaboration spaces MR allows~\cite{herskovitz2022xspace} would be to allow teams to seamlessly transition between individual, sub-group, and collective interaction modes. This would enable groups to balance collective evaluation and individual agency within their problem-solving processes as they deem appropriate. 

A shared agent ensures equal access to information and minimizes communication overhead within the group. Moreover, one person questioning GenAI contributions can prompt a group-wide critical evaluation of it. However, shared agents also risk influencing everyone and fostering echo chambers. Future research should therefore focus on exploring when individual \textit{vs.} collaborative GenAI will be more effective by building on our understanding of group dynamics~\cite{salas2008teams,eloy2023capturing} and social pressures~\cite{kelman2006interests} to study how GenAI influence spreads within groups. This work will be critical to inform the development of new rituals and meeting styles that structure and foster intentional engagement with collaborative GenAI agents during complex problem-solving processes.

\begin{figure} [t]
  \centering
  \includegraphics[width=\linewidth]{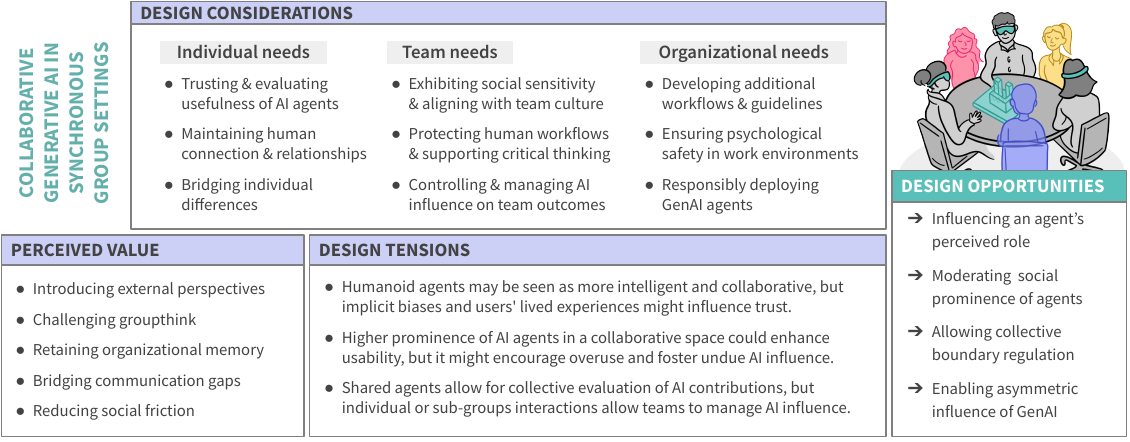}
  \caption{A summary of our insights within this paper that outlines the value end-users see in collaborative use of GenAI and the different design consideration, tensions, and opportunities with respect to it.}
  \Description{}
  \vspace{-1.5em}
  \label{fig:summary}
\end{figure}

\subsubsection*{\textbf{Enabling Asymmetric GenAI Influence: }}
Unlike traditional voice or screen-based interfaces that enable uniform experiences, MR also enables asymmetric collaborative interfaces that allow for nuanced personalization while preserving a shared task space~\cite{johnson2023unmapped,gronbaek2024blended}. Designers can therefore easily tailor collaborative GenAI agents to each team member's role and preferences by adapting agents' perceived role, social presence, and the type of information each individual receives. For example, a system could deliver tailored insights based on team members' expertise and modulate agent representation to shape perceived usefulness for each member. Such a system could theoretically support each individual to contribute to team goals in a way that aligns with their unique role and strengths. This would also allow users to take on different roles that facilitate shared error detection and critical evaluation of GenAI contributions from varied perspectives.

However, adjusting GenAI influence on an individual level can complicate agent influence on team dynamics and outcomes. Differing relationships with shared agents may also lead to variations in reliance and perceived authority that disrupt group cohesion. For instance, individualized spatial presence and customized information access could create disparities in how connected and supported people feel, which in turn could introduce additional power dynamics and communication gaps that weaken common ground. Therefore, while adaptive or asymmetric capabilities can support individual agency, an interesting research direction is exploring how we can leverage individual strengths when interacting with collaborative agents without fragmenting team cohesion. Further research should explore the individual relationships people can form with GenAI agents, and how it affects the maintenance of common ground and the development and evolution of human relationships within the group to fully understand the impact and effectiveness of designing for asymmetric GenAI influence within a team.

%% file: 10_Conclusion.tex
\section{Conclusion}

In this paper, we investigate the potential of \textit{collaborative} GenAI agents to enhance teamwork in synchronous, colocated settings using a speculative approach with a mixed reality \textit{provotype}. We engaged 25 professionals across 6 teams in exploratory workshops and interviews to identify the potential value as well as the individual, team, and organizational user needs that influence acceptance of GenAI agents in synchronous group settings. We also outline how collaborative GenAI systems can support appropriate reliance in groups and highlight the key opportunities spatial and immersive technologies offer to modulate GenAI influence on teams. Our insights are summarized in Figure ~\ref{fig:summary}. Through this work, we hope to provide a nuanced understanding of end-user perspectives and encourage researchers and practitioners to take a careful approach that ensures collaborative GenAI systems truly augment team capabilities without stifling their agency.

%% file: 11_Appendix.tex
\appendix

\section{Provotype Implementation Details} \label{appendix:implementation}

Our provotype implementation integrated an MR application, a GPT-based agent pipeline, and a Wizard-of-Oz interface. 

\begin{figure} [h]
  \centering
  \includegraphics[width=\linewidth]{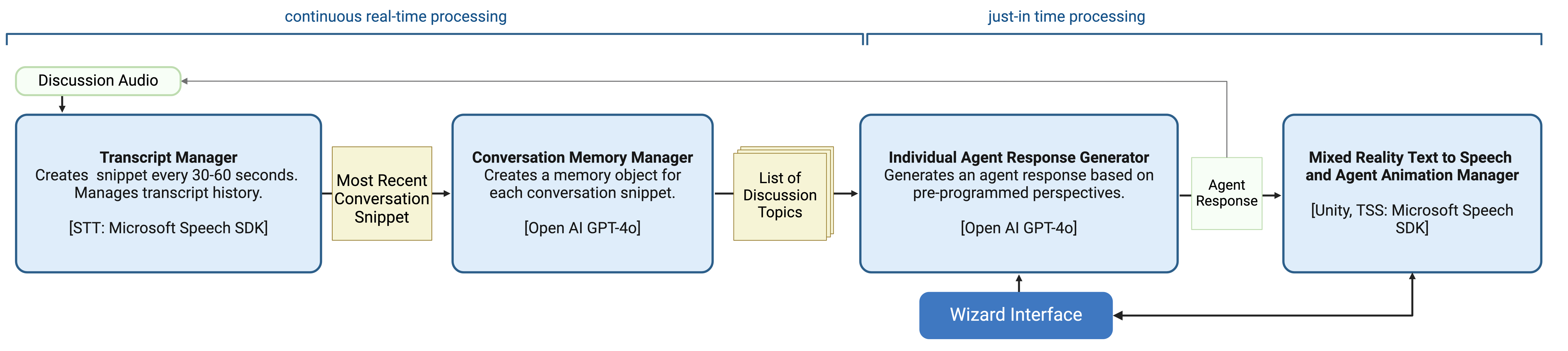}
  \caption{A simplified depiction of our LLM-based pipeline which was used to generate collaborative agent contributions to a group discussion in real-time. It includes a continuous real-time processing component that builds a global conversation memory as the discussion progresses. The current context of the discussion is used to create AI agent opinions, and a just-in-time processing component then creates the agent's response based on both the global memory and current context of the conversation.}
  \Description{}
  \label{fig:pipeline}
\end{figure}

\begin{itemize}
    \item The \textbf{MR application} included a set up scene where researchers could configure and place the GenAI agents in the collaborative space, an welcome scene that participants used to calibrate their headset to their eyes, a collective scene that allowed users to engage in the collective mode, and an individual scene that allowed users to be in the individual mode. The application was built for the Meta Quest Pro headsets using \hyperlink{https://unity.com/}{Unity3D} with Oculus Integration SDK Package~\footnote{https://developers.meta.com/horizon/downloads/package/unity-integration/}. We used the Passthrough API~\footnote{https://developers.meta.com/horizon/documentation/unity/unity-passthrough/} to enable MR, and the Azure AI Speech SDK~\footnote{https://learn.microsoft.com/en-us/azure/ai-services/speech-service/speech-sdk} to handle speech-to-text transcription and text-to-speech synthesis for human-avatar communication. We developed the avatars using the Microsoft Rocketbox Avatar Library~\footnote{https://github.com/microsoft/Microsoft-Rocketbox} and created their rigging and animations with Blender~\footnote{https://www.blender.org/}.

    \item The \textbf{GPT-based agents} were implemented through a multi-stage pipeline to support their ability to contribute to an ongoing group discussion in real-time. This pipeline was built using Python and relied on OpenAI's GPT-4o model~\footnote{https://platform.openai.com/docs/models/gpt-4o} and included a conversation manager, a memory manager and independent modules for each agent for background opinion creation and just-in-time summarization. Details of the pipeline are in Figure~\ref{fig:pipeline}.

    \item The \textbf{Wizard-of-Oz interface} allowed researchers to specify the number of agents that would be integrated into a session and allowed them to control each agents' behavior during the discussion. Each agent had buttons that controlled three behaviors: raise hand, lower hand, and speak. The Wizard could also view agent responses on the interface in real-time. We built this interface using the Javascript and the Jinja web template engine~\footnote{https://jinja.palletsprojects.com/} to manage dynamic page behaviors. Flask~\footnote{https://flask.palletsprojects.com/} was used to handle routing and facilitate communication with the MR application. The web application was deployed to Amazon Web Services (AWS) to allow researchers to access the Wizard-of-Oz interface through any laptop or tablet.
\end{itemize}

\section{Speculative Workshop Material}

\subsection{Concept Videos}
\label{appendix:videos}

\begin{itemize}
    \item Tony Stark interacting with JARVIS from the Marvel series: \href{https://drive.google.com/file/d/1EwMTH68skkGtipvn20a8-iPlAJptoInH/view?usp=drive_link}{Link to video clip}
    
    \item Future Technology Vision Video 1: \href{https://drive.google.com/file/d/19kcQ5YucZuIsVctSR9ReLp1CtNKRSFiO/view?usp=drive_link}{Link to video clip}
    
    \item Future Technology Vision Video 2: \href{https://drive.google.com/file/d/1bpEW6xxmwsfD7iNaEmD7tBW4ic-Ohp6G/view?usp=drive_link}{Link to video clip}
\end{itemize}

\subsection{Workshop Activity Templates} \label{appendix:template}

\begin{figure} [h]
  \centering
  \vspace{-1.25em}
  \begin{minipage}{0.49\textwidth}
  \centering
  \includegraphics[width=\linewidth]{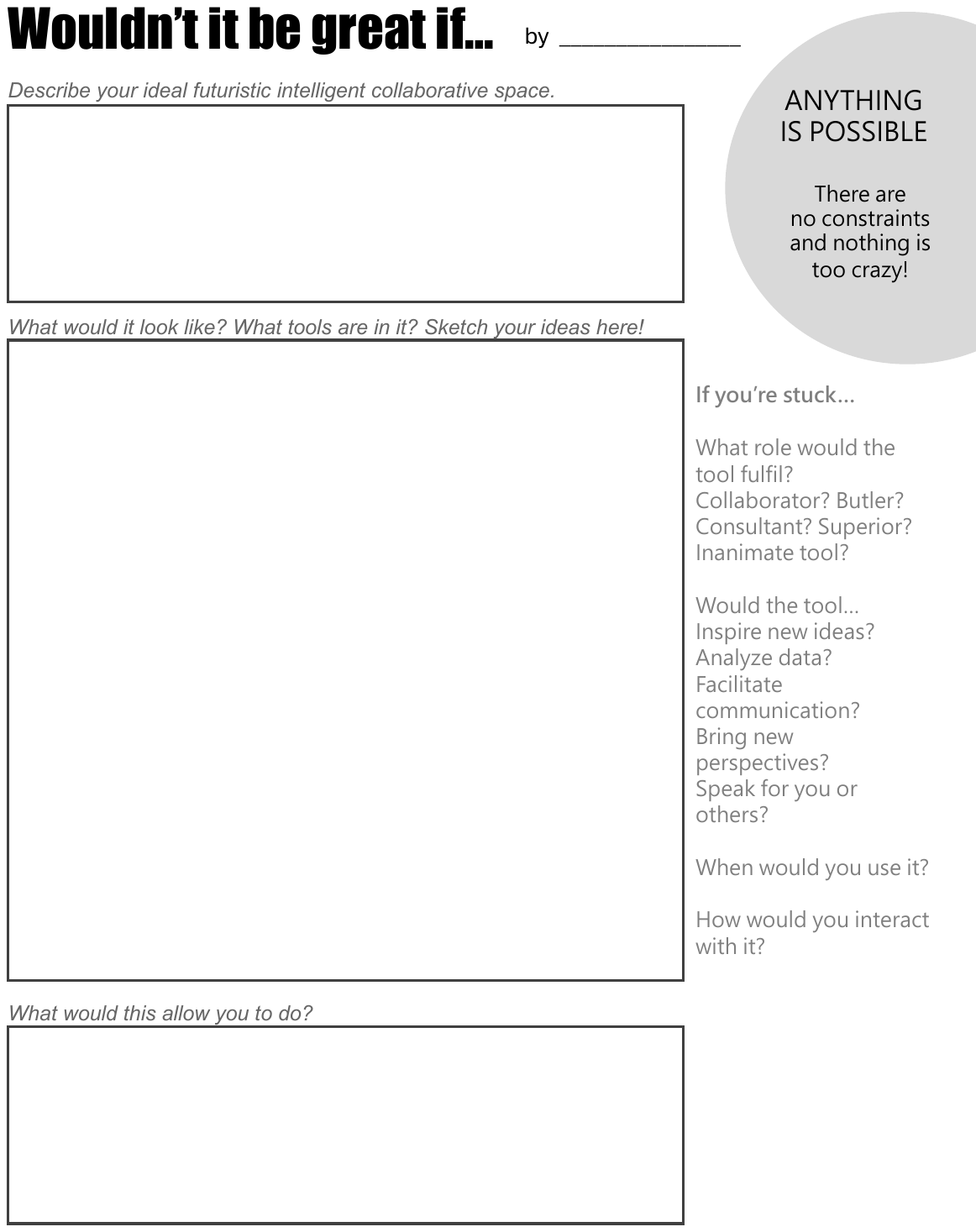}
  \end{minipage}
  \begin{minipage}{0.49\textwidth}
  \centering
  \includegraphics[width=\linewidth]{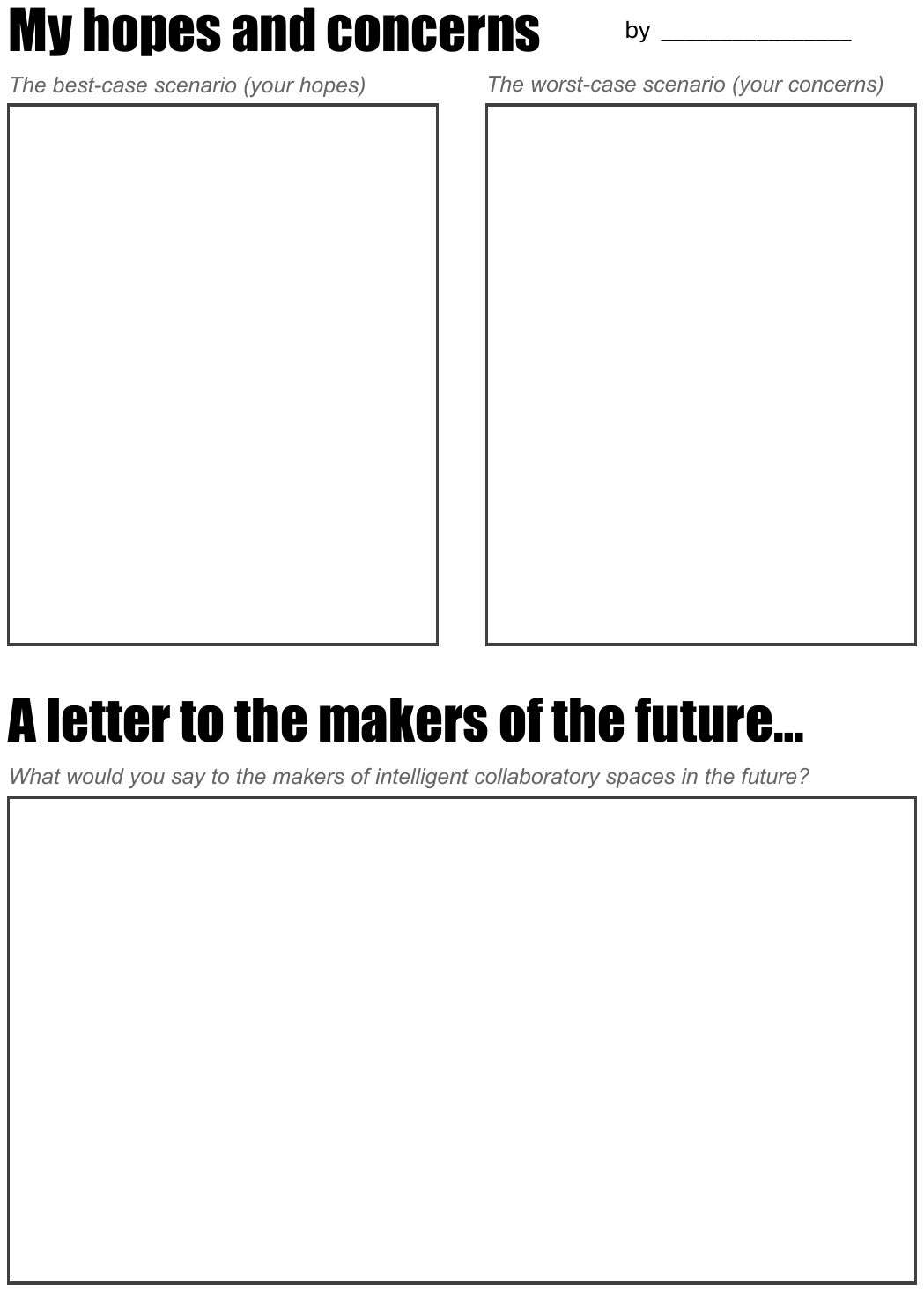}
  \end{minipage}
  \caption{Worksheet templates for the team workshops. Participants used these worksheets to outline their ideal collaborative space (Left), and reflect on their future (Right).}
  \Description{}
  \label{fig:whatiftemplate}
\end{figure}